\documentclass[aps,prd,amssymb,amsmath,preprintnumbers]{revtex4}

\usepackage{graphicx}

\begin{document}

\title{Induced quantum metric fluctuations and the validity of
semiclassical gravity}
\author{B.~L. Hu}
\author{Albert Roura}
\affiliation{Department of Physics, University of Maryland,
College Park, Maryland 20742-4111}
\author{Enric Verdaguer}
\affiliation{Departament de F\'{\i}sica Fonamental and CER en
Astrof\'\i sica, F\'\i sica de Part\'\i cules i Cosmologia,
Universitat de Barcelona, Av.~Diagonal 647, 08028 Barcelona,
Spain}

\begin{abstract}
We propose a criterion for the validity of semiclassical gravity
(SCG) which is based on the stability of the solutions of SCG
with respect to quantum metric fluctuations. We pay special
attention to the two-point quantum correlation functions for the
metric perturbations, which contain both intrinsic and induced
fluctuations.  These fluctuations can be described by the
Einstein-Langevin equation obtained in the framework of
stochastic gravity. Specifically, the Einstein-Langevin equation
yields stochastic correlation functions for the metric
perturbations which agree, to leading order in the large $N$
limit, with the quantum correlation functions of the theory of
gravity interacting with $N$ matter fields.  The homogeneous
solutions of the Einstein-Langevin equation are equivalent to the
solutions of the perturbed semiclassical equation, which describe
the evolution of the expectation value of the quantum metric
perturbations. The information on the intrinsic fluctuations,
which are connected to the initial fluctuations of the metric
perturbations, can also be retrieved entirely from the
homogeneous solutions. However, the induced metric fluctuations
proportional to the noise kernel can only be obtained from the
Einstein-Langevin equation (the inhomogeneous term). These
equations exhibit runaway solutions with exponential
instabilities. A detailed discussion about different methods to
deal with these instabilities is given.  We illustrate our
criterion by showing explicitly that flat space is stable and a
description based on SCG is a valid approximation in that case.
\end{abstract}

\preprint{umdpp 03-67}

\maketitle

\section{Introduction}
\label{sec1}

In this paper we discuss the conditions underlying the validity of
semiclassical gravity (SCG) emphasizing the role of metric
fluctuations induced by the quantum matter sources. SCG is based
on the self-consistent solutions of the semiclassical Einstein
equation for a classical spacetime driven by the expectation
value of the stress tensor operator of quantum matter fields. We
propose a criterion based on stochastic semiclassical gravity
\cite{hu03a,hu03b} and compare it with the recently proposed
criterion by Anderson \emph{et al.} \cite{anderson03} based on
linear response theory. To do this we need to reexamine all the
relevant factors old and new contributing to this issue, such as
the reduction of higher derivative equations, intrinsic and
induced fluctuations, and the relation between stochastic and
quantum correlations.  It also necessitates some clarification of
the relation between our approach based on stochastic dynamics
and the linear response approach, and differences with the
approach pursued by Ford \emph{et al.}
\cite{ford82,kuo93,ford99,ford03,borgman03,ford97,yu99,yu00}
based on the normal-ordering and integration-by-parts procedures
on the stress-energy bitensor.  The connection clarified and the
bridges built in this process are beneficial to further
development of the ``bottom-up'' approaches to quantum gravity
starting from SCG \cite{hu95c,hu96,hu99,hu02}.

\paragraph{Metric fluctuations}

SCG accounts for the averaged back reaction of quantum matter
fields and can be regarded as a mean field approximation that
describes the dynamics of the mean spacetime geometry. However,
it does not account for the effects of the fluctuations of
spacetime geometry, which can also be very important. Consider,
for instance, the metric fluctuations induced by the vacuum
fluctuations of the inflaton field in inflationary cosmological
models. Those fluctuations play a crucial role in the generation
of the primordial inhomogeneities which gave rise to the large
scale structure of the present universe as well as the observed
anisotropies of the cosmic microwave background of radiation.

This paper focuses on the effects of the quantum fluctuations of
the metric. We will restrict our treatment to small metric
perturbations around a given background geometry. (Of course, a
full treatment of those fluctuations would require a complete
theory of quantum gravity). We will linearize and quantize those
metric perturbations including their interaction with the quantum
matter fields. This can be described more precisely in terms of
$N$ identical matter fields. Our approach corresponds then to
computing the quantum correlation functions for the metric
perturbations to leading order in a $1/N$ expansion.

In fact, one can show that the leading order contribution to the
quantum correlation functions in a large $N$ expansion is
equivalent to the stochastic correlation functions obtained in
the context of stochastic semiclassical gravity. Whereas SCG is
based on the semiclassical Einstein equation with sources given
by the expectation value of the stress tensor operator of the
quantum matter fields, stochastic semiclassical gravity is based
on the Einstein-Langevin equation, which has in addition sources
due to the noise kernel.  The noise kernel is the symmetrized
connected part of the two-point quantum correlation function of
the stress tensor operator with respect to the state of the
matter fields, and describes their stress-energy fluctuations.

Making use of the equivalence between quantum and stochastic
correlation functions in stochastic semiclassical gravity, one is
naturally led to separate the symmetrized quantum correlation
function for the metric perturbations (to leading order in $1/N$)
into two separate contributions: the \emph{intrinsic} and the
\emph{induced} fluctuations. The former are connected to the
dispersion of the initial state of the metric perturbations,
whereas the latter are proportional to the noise kernel and are
induced by the quantum fluctuations of the matter fields' stress
tensor operator.

\paragraph{Validity of semiclassical gravity}

Different aspects concerning the validity of the description
provided by SCG in the case of free quantum matter fields in the
Minkowski vacuum state propagating on Minkowski spacetime have
been studied by a number of authors. Most of them considered the
stability of such a solution of SCG with respect to small
perturbations of the metric. Horowitz was the first one to
analyze the equations describing those perturbations, which
involved higher order derivatives (up to fourth order), and found
unstable solutions that grow exponentially with characteristic
timescales comparable to the Planck time
\cite{horowitz80,horowitz81}. This was later reanalyzed by Jordan
with similar conclusions \cite{jordan87a}. However, those unstable
solutions were regarded as an unphysical artifact by Simon, who
argued that they lie beyond the expected domain of validity of
the theory and emphasized that only those solutions which
resulted from truncating perturbative expansions in terms of the
square of the Planck length are acceptable
\cite{simon90,simon91}. Further discussion was provided by
Flanagan and Wald \cite{flanagan96}, who advocated the use of an
order reduction prescription first introduced by Parker and Simon
\cite{parker93} but insisted that even nonperturbative solutions
of the resulting second order equation should be regarded as
acceptable. Following these approaches Minkowski is shown to be a
stable solution of SCG with respect to small metric perturbations.

Anderson, Molina-Par\'\i s and Mottola have recently taken up the
issue of the validity of SCG \cite{anderson03} again. Their starting
point is the fact that the semiclassical Einstein equation will fail
to provide a valid description of the dynamics of the mean spacetime
geometry whenever the higher order radiative corrections to the
effective action, involving loops of gravitons or internal graviton
propagators, become important (see
Refs.~\cite{tsamis96a,tsamis96b,tsamis97,tsamis98} for some attempts
to include those effects).  Next, they argue qualitatively that such
higher order radiative corrections cannot be neglected if the metric
fluctuations grow without bound. Finally, they propose a criterion (a
necessary condition) to characterize the growth of the metric
fluctuations, and hence the validity of SCG, based on the stability of
the solutions of the linearized semiclassical equation.

\paragraph{Our criterion}

In this paper we addresses the issue of the stability of
semiclassical solutions with respect to small quantum
corrections. When the metric perturbations are quantized, the
semiclassical equation can be interpreted as the equation
governing the evolution of the expectation value of the operator
for the metric perturbations. We introduce a stability criterion
based on whether the metric fluctuations grow without bound or
not by considering the behavior of the quantum correlation
functions of the metric perturbations. Furthermore, we emphasize
that one should consider not only the intrinsic fluctuations, but
also the induced ones. In fact, the induced fluctuations play a
crucial role when considering the stability of simple open
quantum systems for several reasons. First, those systems usually
exhibit a characteristic relaxation time so that for much larger
times the contribution from the intrinsic fluctuations becomes
negligible.  Second, after that transient period the stability
around an equilibrium configuration is the result of a balance
between the energy dissipated by the system and the fluctuations
induced by the environment, which is encoded in the so-called
fluctuation-dissipation relation connecting the dissipation and
the noise kernels.

It is true that the effect of intrinsic fluctuations can be
deduced from an analysis of the solutions of the perturbed
semiclassical Einstein equation, but in general one cannot
retrieve the effect of the induced fluctuations from it. This
effect can be properly accounted for in the stochastic
semiclassical gravity framework. Both intrinsic and induced
fluctuations are innate in the Einstein-Langevin equation.

\paragraph{Ford's program}

Ford \cite{ford82} was amongst the first to have noted the
importance of quantum fluctuations in these issues. An earlier
criterion put forth by Kuo and Ford \cite{kuo93} used the
variance of the fluctuations of the stress tensor operator
compared to the mean value as a measure of the validity of SCG.
As pointed out by Hu and Phillips \cite{hu00,phillips00} (see
reply by Ford and Wu \cite{ford03}) such a criterion should be
refined by considering the back reaction of those fluctuations on
the metric.  Ford and collaborators also considered both
intrinsic (``active'') \cite{ford97,yu99,yu00} and induced
(``passive'') \cite{ford82,kuo93,ford99,ford03,borgman03}
fluctuations, but they did not treat them in a unified way and
did not discuss their precise relation to the quantum correlation
function for the metric perturbations. Furthermore, they did not
include the full averaged back reaction of the matter fields
self-consistently, and the contribution from the vacuum
fluctuations in Minkowski space were discarded.  As these issues
have been discussed before by both groups of Ford and Hu, we will
only make a few remarks at the end of this paper.

Here, our attention will be focused on comparing the criteria
based on the linear response approach proposed by Anderson
\emph{et al.}  and our stochastic gravity approach.  Since the
differences in the two ways to address the issue of the validity
of SCG is rooted in the difference between linear response theory
and stochastic dynamics as applied to SCG, we hope that this work
can also serve the purpose of offering a comparison between these
two important approaches exploring the validity of the mean field
approximation.  In the examples provided, we will specialize the
matter fields to the case of free scalar fields, but
generalization to vectorial or fermionic fields should not pose
major difficulty.

\paragraph{Terminology and organization}

To avoid unnecessary ambiguities or confusion in interpretation,
it is useful to clarify the use of some terminology here.

First, a comment on the difference between the \textit{stochastic
gravity program} in general and its present implementation status
in particular.  Stochastic semiclassical gravity can be
understood as the Gaussian approximation to stochastic gravity.
Although technically the actual implementations of stochastic
gravity so far (to which our present discussion applies) have
been restricted to linear metric perturbations around the
background geometry and a Gaussian stochastic source, the
theoretical construct of the stochastic gravity program has a
much broader meaning beyond these limitations. It refers to the
range of theories based on the second and higher order correlation
functions.  Noise can be defined in fully nonlinear theories
(\emph{e.g.} correlation noise \cite{calzetta00a} in the
Schwinger-Dyson equation hierarchy) to some degree \footnote{In
general, it might be necessary to extend the concept of stochastic
process to that of processes with a real and normalized
distribution functional but not necessarily positive definite.},
but one should not expect the simple Langevin form with Gaussian
and additive noise to prevail.  Thus, stochastic gravity in this
broad sense entails the whole hierarchy of correlation functions,
which would imply going beyond order 1/N in the generating
functional.  It could in principle provide the means (similar to
the BBGKY hierarchy in kinetic theory) to access the full theory
of quantum gravity \cite{hu99,hu02}. It is in this sense that we
say stochastic gravity is the intermediate theory between SCG (a
mean field theory based on the expectation value of the energy
momentum tensor of quantum fields) and quantum gravity
(understood as the full hierarchy of correlation functions
retaining complete quantum coherence).

Second, the precise meaning in our use of the terms
\emph{perturbations} and \emph{fluctuations}.  By perturbations
of the metric we mean deviations of the perturbed metric from a
background metric. Perturbations are purely classical and
deterministic in general relativity and SCG. In stochastic
gravity, they are classical but stochastic (with a vanishing
statistical expectation value) so that the background
configuration can be regarded as the expectation value of a
stochastic metric (a complete gauge fixing is required to
meaningfully talk about the expectation value of a metric). In
linear quantum gravity, perturbations are quantum operators. For
a state with a vanishing expectation value, the background metric
can then be regarded as the expectation value of the metric
operator (again a complete gauge fixing is required) times the
identity operator. On the other hand, the term
\emph{fluctuations} is employed only to refer to the statistical
fluctuations of the metric perturbations when they correspond to
a stochastic process, or to the quantum fluctuations of the
metric perturbations when they are treated as a quantum operator.

Third, by \textit{leading order} in the large $N$ limit we mean
the lowest order in $1/N$ with a nonvanishing contribution. Hence,
as we will see,  the leading order for the source of the
semiclassical Einstein equation, which is proportional to the
expectation value of the stress tensor operator, is $1/N^0$,
whereas the leading order for the quantum two-point correlation
functions is $1/N$.

The paper is organized as follows. In Sec.~\ref{sec2} we briefly
review the fundamental aspects of SCG and how one can study
linearized perturbations around a background solution of SCG.
This is generalized to incorporate the metric fluctuations in
Sec.~\ref{sec3}, where the key elements of stochastic
semiclassical gravity are introduced and the equivalence between
stochastic and quantum correlation functions is explained.  In
Sec.~\ref{sec5} we propose a generalized stability criterion that
includes the metric fluctuations, which is then applied to the
specific case of a Minkowski background. We conclude by
summarizing and discussing the main results in Sec.~\ref{sec7}.

A number of additional details and technical points are left for
the Appendices. In Appendix~\ref{appA} we illustrate the basic
aspects of intrinsic and induced fluctuations using a simple
quantum Brownian motion model. In Appendix~\ref{appD} we provide
the expressions for the dissipation and noise kernels in a
Minkowski spacetime and the vacuum state. Some of the main steps
to show the equivalence between stochastic and correlation
functions using a large $N$ expansion are summarized in
Appendix~\ref{appE}. The physical interpretation of the singular
coincidence limit for the noise kernel and possible ways to deal
with it are explained in Appendix~\ref{appB}. Finally, in
Appendix~\ref{appC} we discuss the existence of runaway solutions
in SCG and stochastic semiclassical gravity as well as methods to
deal with them.

Throughout the paper we use natural units with $\hbar=c=1$ and
the $(+,+,+)$ convention of Ref.~\cite{misner73}. We also make use
of the abstract index notation of Ref.~\cite{wald84}. Latin
indices denote abstract indices, whereas Greek indices are
employed when a particular coordinate system is considered.

\section{Semiclassical gravity and linear response theory}
\label{sec2}

A possible first step when addressing the interplay between
gravity and quantum field theory is to consider the evolution of
quantum matter fields (matter field is referred to here as any
field other than the gravitational one) on a classical spacetime
with a nontrivial geometry, characterized by a metric $g_{ab}$.
As opposed to the situation for a Minkowski spacetime, there is
in general no preferred vacuum state for the fields and particle
creation effects naturally arise, such as Hawking radiation for
black holes, cosmological particle creation and the generation of
primordial inhomogeneities in  inflationary cosmological models.
\textbf{Quantum field theory in curved spacetimes} (QFTCST) is by
now a well-established subject (at least for free fields and
globally hyperbolic spacetimes) \cite{birrell94,wald94}.

QFTCST is only an approximation in that the matter fields are
treated as test fields evolving on a given spacetime. Einstein's
theory requires that spacetime dynamics determines and is
determined by the matter field. Thus one  needs to consider the
back reaction of the quantum matter fields on the dynamics of the
spacetime geometry, which naturally leads to the
\textbf{semiclassical theory of gravity}, where the evolution of
the spacetime metric $g_{ab}$ is determined by the semiclassical
Einstein equation
\begin{equation}
G_{ab}[g] + \Lambda g_{ab} - \alpha A_{ab}[g] - \beta B_{ab}[g]
=\kappa \left\langle \hat{T}_{ab}[g] \right\rangle' _\mathrm{ren}
, \label{einstein1}
\end{equation}
where $g_{ab}$ is the spacetime metric, $G_{ab}[g]$ is the
Einstein tensor and the matter source corresponds to the
renormalized expectation value of the stress tensor operator of
the matter fields (a prime was used to distinguish it from that
introduced below after absorbing some terms). Here, $\Lambda$ is
the renormalized cosmological constant, $\kappa=8\pi G$, with $G
\equiv 1/m_p^2$ being the Newton constant and $m_p$ the Planck
mass; $\alpha$ and $\beta$ are renormalized dimensionless
coupling constants associated with tensors $A_{ab}[g], B_{ab}[g]$
needed for the renormalization of the logarithmic divergences
\footnote{The renormalized coupling constants are running
coupling constants which depend on some renormalization scale
$\mu$. However, since $\langle \hat{T}_{ab}[g] \rangle'
_\mathrm{ren}$ has the same dependence on $\mu$, the
semiclassical Einstein equation is invariant under the
renormalization group, which involves changes in the
renormalization scale $\mu$.}. The expectation value of the
stress tensor operator exhibits divergences which are local and
state independent.  Introducing a covariant regularization and
renormalization procedure, those divergences can be absorbed into
the cosmological constant, the Newton constant multiplying the
Einstein-Hilbert term and the gravitational action counterterms
quadratic in the curvature. The finite contributions from those
counterterms give rise to the covariantly conserved tensors
$A_{ab}$ and $B_{ab}$ which result from functionally
differentiating with respect to the metric the terms $\int d^4x
\sqrt{-g}C^{abcd}C_{abcd}$ and $\int d^4x \sqrt{-g} R^2$
respectively, where $C_{abcd}$ is the Weyl tensor and $R$ is the
Ricci scalar. Those contributions were explicitly written on the
left-hand side of Eq.~(\ref{einstein1}), but from now on will be
included in the renormalized expectation value of the stress
tensor operator so that the semiclassical Einstein equation
becomes
\begin{equation}
G_{ab}[g] =\kappa \left\langle \hat{T}_{ab}[g] \right\rangle
_\mathrm{ren} . \label{einstein2}
\end{equation}
The field operators appearing in the stress tensor operator for
the quantum matter fields are in the Heisenberg picture and
satisfy the corresponding equation of motion, which coincides
with the classical field equation for fields evolving on that
spacetime. In particular, if we consider a free scalar field, the
field operator in the Heisenberg picture will satisfy the
corresponding Klein-Gordon equation for that geometry.

Given a manifold $\mathcal{M}$ and a metric $g_{ab}$ which
characterize a globally hyperbolic spacetime, and a density matrix
$\hat{\rho}$ which specifies the state of the quantum matter
fields on a particular Cauchy hypersurface, the triplet
$(\mathcal{M},g_{ab},\hat{\rho})$ constitutes a solution of SCG
if it is a self-consistent solution of both the semiclassical
Einstein equation~(\ref{einstein2}) and the equations of motion
for the quantum operators of the matter fields evolving on the
spacetime manifold $\mathcal{M}$ with metric $g_{ab}$. Those
operators enter in turn into the definition of the stress tensor
operator appearing in the semiclassical Einstein equation.

The semiclassical Einstein equation has been justified in at
least two different ways. One possibility is to argue, by
assuming a number of reasonable axioms, that it is the only
consistent way to extend the classical Einstein equation to
couple the quantum matter fields to a classical metric
\cite{wald94,flanagan96}.  Alternatively, it can be derived by
considering $N$ free matter fields weakly coupled to the
gravitational field in the sense that the gravitational coupling
constant times the number of fields $NG$ remains constant in the
limit $N \rightarrow \infty$ \cite{hartle81}.  The semiclassical
Einstein equation can then be shown to correspond to the
dynamical equation for the evolution of the expectation value of
the metric in the limit of large $N$ \footnote{One could be
concerned that such a derivation was purely formal due to the
impossibility of having a well-defined expectation value for the
metric (at least without a complete gauge fixing), the
difficulties in defining a measure for the path integral free of
problems related to the gauge freedom under diffeomorphisms, and
the issues related to the nonrenormalizable character of
perturbative gravity. Nevertheless, in the limit $N \rightarrow
\infty$ the contribution from the graviton loops vanishes so that
the last two problems become irrelevant, whereas the fluctuations
of the metric around a given background are completely suppressed
and, hence, the first problem also disappears.}. Of course, in
reality $N$ is finite and the semiclassical Einstein equation can
only be understood as the lowest order contribution in a $1/N$
expansion.

Functional methods based on path integrals are useful not only in the
rendition of ideas but also in actual computations. However, the usual
\emph{in-out} formalism suitable for computing transition matrix
elements in scattering problems is not appropriate to deal with
back-reaction problems in which one is interested in the causal
evolution of true expectation values from their initial values. The
closed-time-path (CTP) formalism, which naturally yields true
expectation values and causal evolution equations for their dynamics
\cite{schwinger61,bakshi63,keldysh65,chou85,jordan86}, should be used
instead. This formalism has been applied to study a number of
situations involving gravitational back-reaction effects of quantum
fields \cite{calzetta87,campos94,calzetta94,hu95a,calzetta97c}.  (See
also Ref.~\cite{cooper94} for an interesting application to an
analogous situation in QED, where the back reaction of charged quantum
fields on the dynamics of the expectation value of the electromagnetic
field was considered.) This includes
minisuperspace models which restrict the possible geometries to
Robertson-Walker metrics and  consider perturbative deviations
from the massless conformal case for the matter fields
\cite{calzetta94,hu95a,calzetta97c}, as well as
small metric perturbations (of a less restricted form) around
self-consistent solutions of SCG \cite{calzetta87,campos94}.

More specifically, given a background metric $g_{ab}$ which is a
solution of the semiclassical Einstein equation in SCG, one can
compute the CTP effective action on the perturbed metric
$\tilde{g}_{ab}=g_{ab}+h_{ab}$ up to quadratic order in the metric
perturbations $h_{ab}$. Taking the functional derivative of the
CTP effective action with respect to $ h_{ab} $, one gets the
perturbed version of Eq.~(\ref{einstein2}) to linear order in the
metric perturbations:
\begin{equation}
G_{ab}^{(1)}\left[ g+h\right] = \kappa \left\langle
\hat{T}_{ab}^{(1)} [g+h] \right\rangle _\mathrm{ren}
\label{einstein4},
\end{equation}
where the superindex $(1)$ was used to denote that only terms
linear in $h_{ab}$ should be considered. The linearized Einstein
tensor $G_{ab}^{(1)} [g+h]$ comes from the Einstein-Hilbert term
in the gravitational action. On the other hand, the contribution
to the CTP effective action which results from functionally
integrating the matter fields involving the expectation value and
two-point correlation functions of the stress tensor operator on
the background geometry yields the following result for the
linearized expectation value of the stress tensor operator:
\begin{equation}
\left\langle \hat{T}^{(1)\, ab}[g+h;x) \right\rangle = - 2
\left(H \cdot h\right)^{ab}(x) - 2\left(M \cdot h\right)^{ab}(x)
\label{STexpecvalue},
\end{equation}
where we have introduced the notation $A\cdot B\equiv \int d^{4}y
\sqrt{-g(y)} A^{ab}(y) B_{ab}(y)$, and the kernels $H$ and $M$ are
given by
\begin{eqnarray}
H^{abcd}\left( x,y\right)  &=& -\frac{1}{4} \mathrm{Im}
\left\langle T^{*}\hat{T}^{ab}\left[ \hat{\varphi},g;x\right)
\hat{T}^{cd}\left[ \hat{\varphi},g;y\right) \right\rangle
+\frac{i}{8}\left\langle \left[ \hat{T}^{ab}
\left[ \hat{\varphi},g;x\right) ,\hat{T}^{cd}\left[
\hat{\varphi},g;y\right) \right] \right\rangle
\label{Hkernel} \\
M^{abcd}(x,y) &=& -\frac{1}{2}\left( \frac{1}{\sqrt{-g(x)}}
\frac{\delta \left(\left\langle
\hat{T}^{ab}[\hat{\varphi},g_{ab};x) \right\rangle
\right)}{\delta g_{cd}(y)} \right) ,  \label{Mkernel}
\end{eqnarray}
where the notation $T^*$ was employed to indicate that the
spacetime partial derivatives appearing in the time-ordered
operators also act on the theta function implementing the time
ordering. The functional derivative appearing on the right-hand
side of Eq.~(\ref{Mkernel}) should be understood to account only
for the explicit dependence on the metric: the implicit dependence
through the field operator $\hat{\varphi}[g]$ is excluded
\footnote{The kernels $H$ and $M$ both exhibit divergences, but
they can be removed by the standard procedure for renormalizing
the expectation value of the stress tensor in an arbitrary
spacetime (after all they are related to the terms in $\langle
\hat{T}^{ab}[g+h;x) \rangle$ that are linear in $h_{ab}$), which
involves renormalizing the cosmological constant and the Newton
constant in the bare gravitational action as well as including
countertems quadratic in the curvature. More precisely, by
evaluating all the counterterms in the bare gravitational action
on the perturbed metric and keeping the terms quadratic in the
metric perturbations, which give rise to linear terms in $\langle
\hat{T}^{ab}[g;x) \rangle$, the divergences in the CTP effective
action arising from the kernels $H$ and $M$ are exactly
canceled.}.

The previous result for the expectation value of the stress tensor
when small metric perturbations around a background solution of SCG
are considered has been obtained in two ways: 1) by applying the usual
techniques of linear response theory (see, for instance,
\cite{kubo66}) to SCG \cite{mottola86,campos98,anderson03}, and 2) by
employing the influence functional formalism
\cite{feynman63,feynman65} to linearized metric perturbations regarded
as an open quantum system interacting with an environment constituted
by the quantum matter fields ~\cite{martin99b,martin99c}. The
influence functional approach also provides the noise kernel which
underscores the stochastic nature of the dynamics for the metric
perturbations.  We will employ this method in the next Section.

The explicit expression for the linearized expectation value
$\langle \hat{T}^{(1)\, ab}[g+h;x) \rangle$ in the particular
case of a Minkowski background spacetime and a free scalar field
in the Minkowski vacuum state was obtained in
Refs.~\cite{horowitz80} and \cite{flanagan96} for a massless
field and  in Refs.~\cite{flanagan98,martin00,anderson03} for a
field with an arbitrary mass. For our discussion we have included
them in Appendix~\ref{appD}, where a global inertial coordinate
system $\{x^\mu\}$ for the Minkowski background is used.
According to Eq.~(\ref{STexpecvalue2}), $\langle \hat{T}^{(1)\,
ab} [g+h] \rangle _\mathrm{ren}$ can be written entirely in terms
of the linearized Einstein tensor $G^{(1)\, ab}$. Taking that into
account, the expression for the linearized semiclassical Einstein
equation~(\ref{einstein4}) in Fourier space becomes
\begin{equation}
F^{\mu\nu}_{\hspace{2ex} \alpha\beta}(p) \, \tilde{G}^{(1)\,
\alpha\beta}(p) = 0 \label{einstein3},
\end{equation}
where
\begin{equation}
F^{\mu\nu}_{\hspace{2ex} \alpha\beta}(p) = F_1(p) \, \delta^\mu_{(
\alpha} \delta^\nu_{\beta )}+ F_2(p) \, p^2 P^{\mu\nu}
\eta_{\alpha\beta} \label{Fkernel},
\end{equation}
with
\begin{eqnarray}
F_1(p) = 1 + 2\kappa p^2 \left[ \tilde{H}_\mathrm{A}(p)
- 2 \bar{\alpha} \right] \label{F1},\\
F_2(p) = -\frac{2\kappa}{3} \left[ \tilde{H}_{\rm A}(p) + 3
\tilde{H}_{\rm B}(p) -2 \bar{\alpha}-6 \bar{\beta}\right]
 \label{F2}.
\end{eqnarray}
where $\bar{\alpha}$ $\bar{\beta}$ are some constants which
include the renormalized parameters $\alpha$ and $\beta$ in
Eq.~(\ref{einstein1}), and the kernels $\tilde{H}_A(p)$ and
$\tilde{H}_B(p)$ are defined in Eqs.~(\ref{HAkernel}) and
(\ref{HBkernel}) of Appendix~\ref{appD}. The solutions of
Eq.~(\ref{einstein3}) were analyzed in Refs.~\cite{horowitz80} and
\cite{flanagan96} for the massless case and Ref.~\cite{anderson03}
for the general case. There is an obvious solution for
$\tilde{G}^{(1)}_{ab}(p)=0$, which corresponds to the linear
gravitational waves propagating in Minkowski spacetime. In
addition, there are solutions of the form
$\tilde{G}^{(1)}_{\mu\nu} \propto \delta(p^2-p_0^2)$ for
particular values of $p_0^2$ (positive or negative) comparable to
$m_p^2$. Since they exhibit characteristic timescales or
lengthscales comparable to the Planck scale, where semiclassical
gravity is not expected to be reliable anymore, they are usually
regarded as unphysical. A more detailed discussion on this kind
of solutions is given in Sec.~\ref{sec5} and Appendix~\ref{appC}.

In fact, as will be explained in Sec.~\ref{sec5}, if one
quantizes the linearized metric perturbations,
Eq.~(\ref{einstein3}) coincides with the equation governing the
evolution of the expectation value of the operator $\hat{h}_{ab}$
for the metric perturbations. Therefore, an analysis of the
stability of the solutions of linearized semiclassical Einstein
equation (\ref{einstein3}) can be equivalently understood in
terms of the evolution for the expectation value of
$\hat{h}_{ab}$. In Sec.~\ref{sec5} we will argue that a stability
analysis for solutions of SCG with respect to small quantum
corrections based solely on the expectation value of the metric
perturbations is incomplete and should be extended to take into
account the metric fluctuations as well. Stochastic semiclassical
gravity is particularly well-suited to study the fluctuations of
the metric and will play an important role in our later
discussions. Therefore, in the next section we briefly review the
formalism based on the Einstein-Langevin equation for small
metric perturbations around semiclassical solutions within the
framework of stochastic semiclassical gravity.

\section{Einstein-Langevin equation for metric perturbations around a
given background} \label{sec3}

The semiclassical Einstein equation, which takes into account only
the mean values, is inadequate whenever the fluctuations of the
stress tensor operator are important. An improved treatment is
provided by the Einstein-Langevin equation of \textbf{stochastic
gravity}, which contains a (Gaussian) stochastic source with a
vanishing expectation value and a correlation function
characterized by the symmetrized two-point function of the stress
tensor operator. This theory has been discussed by a number of
authors
\cite{calzetta94,hu95a,hu95b,campos96,calzetta97c,martin99b,hu03a,hu03b}.
Consider a globally hyperbolic background spacetime and an
initial state for the quantum matter fields (one usually
restricts to free fields) which is a self-consistent solution of
SCG, \emph{i.e.}, it satisfies the semiclassical Einstein
equation with the expectation value of the stress tensor operator
obtained by considering the evolution of the matter fields on the
same background geometry. The Einstein-Langevin equation
governing the dynamics of the linearized perturbations $h_{ab}$
around the background metric $g_{ab}$ is given by
\begin{equation}
G_{ab}^{(1)}\left[ g+h\right] =\kappa \left\langle
\hat{T}_{ab}^{(1)} [g+h] \right\rangle _\mathrm{ren}+\kappa \,\xi
_{ab}\left[ g\right] , \label{langevin}
\end{equation}
where the Gaussian stochastic source $\xi _{ab}[g]$ is completely
characterized by its correlation function in terms of the noise
kernel $\mathcal{N}_{abcd}(x,y)$, which accounts for the
fluctuations of the stress tensor operator, as follows:
\begin{equation}
\left\langle \xi _{ab}[g;x) \xi _{cd}[g;y) \right\rangle _{\xi } =
\mathcal{N}_{abcd}(x,y) \equiv \frac{1}{2}\left\langle
\left\{\hat{t}_{ab}[g;x), \hat{t}_{cd}[g;y) \right\}
\right\rangle , \label{noise}
\end{equation}
where $\hat{t}_{ab} \equiv
\hat{T}_{ab}-\langle\hat{T}_{ab}\rangle$ and $\langle \ldots
\rangle$ is the usual expectation value with respect to the
quantum state of the matter fields, whereas $\langle \ldots
\rangle_\xi$ denotes taking the average with respect to all
possible realizations of the stochastic source $\xi_{ab}$. Note
that any local term quadratic in the curvature arising from
finite contributions of the counterterms required to renormalize
the bare expectation value of the stress tensor operator has been
absorbed into its renormalized version $\langle \hat{T}_{ab}^{(1)}
[g+h] \rangle_\mathrm{ren}$.  It should also be emphasized that
solutions of the Einstein-Langevin equation for the metric
perturbations are classical \emph{stochastic} tensorial fields,
not  quantum operators.

The precise meaning that should be given to these stochastic
metric perturbations and the relation of the corresponding
stochastic correlation functions to the quantum fluctuations
resulting from quantizing these metric perturbations will be
discussed below. Before doing so, it is, however, useful to
mention some of the basic properties of the Einstein-Langevin
equation (a more detailed discussion can be found in
Refs.~\cite{martin99b,hu03a,hu03b}). First, when taking the
average of the Einstein-Langevin equation (\ref{langevin}) with
respect to all the possible realizations of the stochastic
source, we recover the semiclassical Einstein
equation~(\ref{einstein4}), as follows straightforwardly from the
vanishing expectation value of the stochastic source. Secondly,
the integrability of the Einstein-Langevin equation is
guaranteed, in the same way as in the semiclassical Einstein
equation, by the conservation of the matter sources. The
conservation of the expectation value of the stress tensor
operator follows immediately from the fact that the divergence
(with respect to the covariant derivative) of the stress tensor
operator vanishes when the equation of motion of the matter field
operators is satisfied (\emph{e.g.} the Klein-Gordon equation for
a scalar field). On the other hand, the fact that
$\langle\nabla^a\xi_{ab}(x)\rangle_\xi=0$ and
$\langle\xi_{ab}(x)\nabla^c\xi_{cd}(x)\rangle_\xi=0$, which
completely characterize the Gaussian stochastic field
$\nabla^a\xi_{ab}(x)$, guarantees the conservation of the
stochastic source (unless otherwise stated, from now on all the
covariant derivatives are taken in the background metric and
indices are raised and lowered using also the background metric).
The previous two equalities are, respectively, a consequence of
the vanishing expectation value of the stochastic source and the
fact that $\nabla^a\hat{t}_{ab}(x)=0$. Finally, the
Einstein-Langevin equation is invariant under gauge
transformations corresponding to infinitesimal diffeomorphisms
characterized by an arbitrary vector field $\vec{\zeta}(x)$,
which generate transformations for the metric of the form $h_{ab}
\rightarrow h_{ab} + \nabla_a\zeta_b + \nabla_b\zeta_a$. This
fact can be seen by realizing that the stochastic source does not
depend on the metric perturbations, whereas the terms depending
on the metric perturbations are all together gauge invariant.
This is because they correspond to perturbing the semiclassical
Einstein equation for the background metric $g_{ab}$, which is
automatically satisfied since the background configuration under
consideration for the metric and the state of the matter fields
is a solution of SCG.

The Einstein-Langevin equation had been previously derived making
use of a formal analogy with open quantum systems and employing
the influence functional formalism \cite{feynman63,feynman65}.
This form was also justified in Ref.~\cite{martin99a} by arguing
that it is the only consistent generalization of the semiclassical
Einstein equation which takes into account the lowest order
effects due to the fluctuations of the stress tensor operator. In
fact, making use of a large $N$ expansion, one can show that the
stochastic correlation functions for the metric perturbations
obtained from the Einstein-Langevin equation coincide with the
leading order contribution to the quantum correlation functions
in the large $N$ limit. The details of the derivation will be
given in Ref.~\cite{roura03b} and are summarized in
Appendix~\ref{appE} for the particular case of a Minkowski
background, to which we will restrict in the present discussion.
In particular, the two-point stochastic correlation function is
equivalent to the symmetrized quantum correlation function to
leading order in $1/N$ provided that one also averages over the
initial conditions for the solutions of the Einstein-Langevin
equation distributed according to the Wigner functional
characterizing the initial state of the metric perturbations (see
Eq.~(\ref{wigner}) in Appendix~\ref{appE} for the definition of
the Wigner functional). It is, therefore, convenient to express
the solutions of the Einstein-Langevin equation as
\begin{equation}
h_{ab} (x) = \Sigma_{ab}^{(0)}(x) + \bar{\kappa} (G_\mathrm{ret}
\cdot \xi)_{ab} (x) \label{solution2},
\end{equation}
where $\bar{\kappa} = N \kappa$ is the rescaled gravitational
coupling constant introduced in Appendix~\ref{appE},
$\Sigma_{ab}^{(0)}(x)$ is a solution of the homogeneous part of
the Einstein-Langevin equation~(\ref{langevin}) containing all the
information about the initial conditions (by homogeneous part we
mean Eq.~(\ref{langevin}) excluding the stochastic source, which
coincides with the semiclassical Einstein
equation~(\ref{einstein2})), and $G_\mathrm{ret} (x,x')$ is the
retarded propagator with vanishing initial conditions associated
with that equation (see Sec.~\ref{appC.3} in Appendix~\ref{appC}
for important remarks on the propagator). Using
Eq.~(\ref{noise}), we can then get the following result for the
symmetrized two-point quantum correlation function:
\begin{equation}
\frac{1}{2}\left\langle \left\{ \hat{h}_{ab}(x), \hat{h}_{cd}(x')
\right\} \right\rangle = \left\langle \Sigma_{ab}^{(0)}(x)
\Sigma_{cd}^{(0)}(x') \right\rangle _{\Sigma_{ab}^{(i)},
\Pi^{cd}_{(i)}} + \frac{\bar{\kappa}^2}{N} \left( G_\mathrm{ret}
\cdot \mathcal{N} \cdot (G_\mathrm{ret})^{T} \right)_{abcd}
(x,x') \label{correlation2},
\end{equation}
where the Lorentz gauge condition $\nabla^a (\bar{h}_{ab} - 1/2
\eta_{ab} h^a_a) = 0$ as well as some initial condition to fix
completely the remaining gauge freedom of the initial state
should be implicitly understood, and the stochastic source was
rescaled according to Appendix~\ref{appE} so that $\langle \xi
_{ab}[g;x) \xi _{cd}[g;y) \rangle_{\xi } = (1/N)
\mathcal{N}_{abcd}(x,y)$, where $\mathcal{N}_{abcd}(x,y)$ is the
noise kernel for a single field.

This result is analogous to that obtained in
Ref.~\cite{calzetta03a} for linear QBM models and briefly
summarized in Appendix~\ref{appA}. It should be emphasized that,
similar to that case, there are two different contributions to
the symmetrized quantum correlation function. The first one is
connected to the quantum fluctuations of the initial state of the
metric perturbations and we will refer to it as \emph{intrinsic
fluctuations}. The second contribution, proportional to the noise
kernel, accounts for the fluctuations due to the interaction with
the matter fields, and we will refer to it as \emph{induced
fluctuations}. In the next section we will formulate a generalized
stability criterion for the solutions of SCG which involves the
quantum fluctuations of the metric. In particular we will see
that the induced fluctuations will play an important role on that
issue.


The noise kernel that we need for our discussions is for the
particular case of a Minkowski background spacetime with a scalar
field in the Minkowski vacuum. It  was obtained in
Ref.~\cite{martin00} and is given by Eq.~(\ref{noiseMink}) in
Appendix~\ref{appD}.

\section{Stability criterion for solutions of semiclassical gravity}
\label{sec5}

In this Section we will propose a criterion for analyzing the
stability of a given solution of SCG with respect to small
quantum corrections, associated with quantized metric
perturbations around a background geometry. As an important
example, we will apply this to the particular case of a Minkowski
background with $N$ scalar fields in the Minkowski vacuum state.

\subsection{Stability of Minkowski space: previous criteria}
\label{sec5.1}

The stability of metric perturbations around a Minkowski spacetime
interacting with quantum matter fields in their Minkowski vacuum
state was first studied in the context of SCG by Horowitz
\cite{horowitz80}. He considered massless conformally coupled
scalar fields and found exponential instabilities for the
linearized metric perturbations with characteristic timescales
comparable to the Planck time. Those solutions are closely
related to the higher derivative countertems required to
renormalize the expectation value of the stress tensor operator
(see, however, Appendix~\ref{appC} for further comments on this
point) and are analogous to the runaway solutions commonly
present in radiation reaction processes such as those considered
in classical electrodynamics \cite{jackson99,johnson02}. It is
generally believed that the runaway solutions obtained by
Horowitz are an unphysical artifact since they involve scales
beyond the regime where SCG is expected to be reliable (in fact,
this statement can be naturally formulated when regarding general
relativity as a low energy effective theory).

Since the existence of terms with higher derivatives in time
implies an increase in the number of degrees of freedom (in an
initial value formulation, not only the metric and its time
derivative should be specified, but also its second and third
order time derivatives), it seems plausible that, by restricting
to an appropriate subspace of solutions of the semiclassical
Einstein equation, one can reestablish the usual number of
degrees of freedom in general relativity and, at the same time,
get rid of all the unphysical runaway solutions. Following this
line of thought Simon proposed that one should restrict to
solutions which result from truncating to order $\hbar$ an
analytic expansion in $\hbar$ (or equivalently in $l_p^2$, the
Planck length squared) \cite{simon90,simon91}. Together with
Parker he also introduced a prescription to reduce the order of
the semiclassical Einstein equation which was computationally
convenient in order to obtain solutions corresponding to such
truncated perturbative expansions in $\hbar$ \cite{parker93}.

On the other hand, Flanagan and Wald argued that Simon's criterion
based on truncating to order $\hbar$ solutions which correspond to
analytic expansions in $\hbar$ seemed too restrictive since it
only allowed small deviations with respect to the classical
solutions of the Einstein equations \cite{flanagan96}. In
particular, one would miss those situations in which the small
semiclassical corrections build up to give significant deviations
at long times, such as those corresponding to the evaporation of
a macroscopic black hole (with a mass much larger than the Planck
mass) by emission of Hawking radiation. Furthermore, they
illustrated with simple examples that there are cases in which
one expects that no solutions of the semiclassical equation are
analytic in $\hbar$. Therefore, they suggested that, rather than
trying to restrict the subspace of acceptable solutions, one
should simply transform the semiclassical equation, by making use
of Simon and Parker's order reduction prescription, to a second
order equation which were equivalent to the original equation up
to the order in $\hbar$ (or $l_p^2$) under consideration. All the
solutions of the second order equation should then be regarded as
acceptable, even if they are not analytic in $\hbar$. Obviously,
one could only extract physically reliable information from those
solutions for scales much larger than the Planck length.

Yet another prescription was proposed by  Anderson,
Molina-Par\'\i s and Mottola \cite{anderson03} on the stability
of small metric perturbations around the Minkowski spacetime.
They got rid of the unphysical runaway solutions by working in
Fourier space and discarding those solutions which corresponded
to 4-momenta with modulus comparable or larger in absolute value
than the Planck mass. However, it is not clear how this procedure
could be generalized to situations where working in Fourier space
is not adequate, as in  time-dependent background spacetimes.

The consequences of both the order reduction prescription
introduced by Simon and Parker and advocated by Flanagan and
Wald, and the procedure employed by Anderson \emph{et al.} are
rather drastic, at least when applied to the case of a Minkowski
background, since one is just left with the solutions of the
sourceless classical Einstein equation corresponding to linear
gravitational waves propagating in Minkowski. In fact, the
situation was not completely trivial for Flanagan and Wald, who
were interested in analyzing whether the averaged null energy
condition (ANEC) was satisfied in SCG by considering
perturbations of the Minkowski solution, because they also
perturbed the state of the matter fields. The order reduction
prescription also seems to exclude those solutions which
correspond to inflationary models driven entirely by the vacuum
polarization of the quantum matter fields \cite{simon92}, such as
the trace anomaly driven inflationary model initially proposed by
Starobinsky \cite{starobinsky80}. To keep this kind of models,
Hawking, Hertog and Reall considered a less drastic alternative
to deal with the runaway solutions \cite{hawking01,hawking02}.
Their procedure, which is analogous to some methods previously
employed in classical electrodynamics for radiation reaction
problems \cite{jackson99}, is based on discarding solutions which
grow without bound at late times (see Appendix~\ref{appC} for
further discussions on this and related issues).


\subsection{Generalized stability criterion}
\label{sec5.2}

How does one characterize the quantum state of the metric
perturbations? The first candidate is the expectation value for
the operator associated with the perturbation of the metric,
$\hat{h}_{ab}$. In fact, using a large $N$ expansion, Hartle and
Horowitz showed that the semiclassical Einstein equation can be
interpreted as the equation governing the evolution of the
expectation value of the metric to leading order in $1/N$
\cite{hartle81}.  Taking that result into account, the study of
the stability of a solution of SCG by
linearizing the semiclassical Einstein equation with respect to
small metric perturbations around that solution can be understood
in the following way: Take an initial state for the metric
perturbations with a small nonvanishing expectation value for the
operator $\hat{h}_{ab}$, let it evolve, and see if  the
expectation value grows without bound.

However, in addition to the expectation value of $\hat{h}_{ab}$
the state of the metric perturbations will also be characterized
by its fluctuations. In fact, if there was no interaction with
matter fields so that the state for the metric perturbations
evolved unitarily, the set of quantum correlation functions (for
the operator $\hat{h}_{ab}$) evaluated at equal times would
completely characterize the quantum state of the metric
perturbations \footnote{Nevertheless, since the metric
perturbations constitute an open quantum system due to the
interaction with the matter fields, their state should be
described by a density matrix (the reduced density matrix
obtained by taking the density matrix for the whole system
--metric perturbations plus matter fields-- and tracing out the
matter fields) which exhibits a nonunitary and even non-Markovian
evolution. Therefore, as explained in Ref.~\cite{calzetta03a},
the correlation functions involving different times may contain
information which cannot be obtained just from the correlation
functions evaluated at equal times.}. Let us now suppose that the
evolution of the expectation value is stable (\emph{i.e.}, that
it does not grow unboundedly with time) or even that it vanishes
for all times. It is clear that the semiclassical solution cannot
be regarded as stable with respect to small quantum corrections
if the fluctuations of the state for the metric perturbations
grow without bound. Therefore, the stability criterion stated in
Ref.~\cite{anderson03} should be generalized: one also needs to
take into account the fluctuations. According to
Ref.~\cite{anderson03}, a necessary condition for the stability
of a solution of SCG requires that no gauge
invariant scalar quantity constructed just from the linearized
metric perturbation $h_{ab}$ (which satisfies the semiclassical
Einstein equation linearized around the semiclassical solution
under consideration) and its derivatives grows without bound.
This criterion can be interpreted as a condition on the stability
of the expectation value of the operator $\hat{h}_{ab}$ for the
state of the metric perturbations. We claim that, in addition,
the $n$-point quantum correlation functions for the metric
perturbations (starting with $n=2$) should also be stable.
Considerations based on gauge-invariant variables will not be
necessary because we will be dealing with expressions where the
gauge freedom has been completely fixed.

As explained in Appendix~\ref{appE}, to leading order in $1/N$ the CTP
generating functional for the metric perturbations exhibits a Gaussian
form provided that a Gaussian initial state for the metric
perturbations with vanishing expectation value is chosen. All the
$n$-point quantum correlation functions can then be obtained, to
leading order in $1/N$, from the two-point quantum correlation
function. Furthermore, any of the two-point quantum correlation
functions can in turn be expressed in terms of the symmetrized and
antisymmetrized correlation functions (the expectation values of the
commutator and anticommutator of the operator $\hat{h}_{ab}$). To
leading order in $1/N$ the commutator is independent of the initial
state of the metric perturbations and is given by
$2i\kappa (G_\mathrm{ret}(x',x) - G_\mathrm{ret}(x,x'))$. On the
other hand, the expectation
value of the anticommutator is given by Eq.~(\ref{correlation2}) and
is the sum of two separate contributions: the intrinsic and the
induced fluctuations.

The first contribution in Eq.~(\ref{correlation2}) to the correlation
function for the metric perturbations involves the solutions of the
homogeneous part of the Einstein-Langevin equation~(\ref{langevin}),
which actually coincides with the linearized semiclassical equation for
the metric perturbations around the background geometry. Similarly,
$G_\mathrm{ret}$ corresponds to the retarded propagator (with
vanishing initial conditions) associated with the linearized semiclassical
equation. Thus, solving the perturbed semiclassical Einstein equation
not only accounts for the evolution of the expectation value of the metric
perturbations, which will exhibit a nontrivial dynamics as long as we
choose an initial state with nonvanishing expectation value, but also
provides nontrivial information, even for a state with a vanishing
expectation value, about the commutator as well as the intrinsic
fluctuations of the metric. This implies that the analysis about the
stability of the solutions of SCG can also be used to
determine the stability of the metric perturbations with respect to
intrinsic fluctuations.

The new observation we make here is that the induced fluctuations can
be important as well. Both the retarded propagator and the solutions
of the linearized semiclassical Einstein equation depend, through the
kernel $H$, on the expectation value of the commutator of the stress
tensor operator on the background geometry and on the imaginary part
of its time-ordered two-point function. However, they do not involve
the expectation value of the anticommutator, which drives the induced
fluctuations. Furthermore, although the expectation value of the
commutator and the anticommutator are related by a
fluctuation-dissipation relation in some particular cases
\cite{martin99b,martin00}, that is not true in general and the induced
fluctuations need to be explicitly analyzed.

To sum up, when analyzing the stability of a solution of SCG
with respect to small quantum corrections, one should also
consider the behavior of both the intrinsic and induced
fluctuations of the quantized metric perturbations. Whereas
information on the stability of the intrinsic fluctuations can be
retrieved from an analysis of the solutions of the perturbed
semiclassical Einstein equation, the effect of the induced
fluctuations is properly accounted for only in the stochastic
semiclassical gravity framework based on the Einstein-Langevin
equation.

\subsection{Stability of Minkowski space from our criterion}
\label{sec6}

We now turn to the application of the criterion proposed in the
previous subsection to the particular yet important case of
Minkowski spacetime. As explained there, the existing results in
the literature can be interpreted as analysis of the stability of
the expectation value of the operator associated with the metric
perturbations (see, however,
Refs.~\cite{hartle81,horowitz81,jaekel95}). On the other hand, we
also need to include in our consideration the fluctuations,
characterized by the two-point quantum correlation function.

Before proceeding to analyze the two-point quantum correlation functions
it is convenient to decompose the metric perturbations around
Minkowski in the following way \cite{anderson03}:
\begin{equation}
h_{ab} = \phi\,\eta_{ab} + \left( \nabla_{(a}\nabla_{b)} - \eta_{ab} \Box
\right) \psi + 2 \nabla_{(a} v_{b)} + h_{ab}^\mathrm{TT},
\end{equation}
where $v^a$ is a transverse vector and $h_{ab}^\mathrm{TT}$ is a
transverse and traceless symmetric tensor, \emph{i.e.}, $\nabla_a
v^a=0$, $\nabla^a h_{ab}^\mathrm{TT}=0$ and $(h^\mathrm{TT})^a_a=0$.
Similarly, any vector field $\zeta^a$ characterizing an infinitesimal
gauge transformation can be decomposed as $\zeta^a = \nabla^a \zeta +
V^a$, where $V^a$ is a transverse vector field. It is then clear that
the vectorial and one of the scalar parts of the metric perturbation
corresponding to $v^a$ and $\psi$ respectively, can be eliminated by
choosing a gauge transformation such that $V^a=-v^a$ and
$\zeta=-\psi/2$ (this will also imply a change for $\phi$: $\phi
\rightarrow \phi + 2 \Box \zeta$).

When the Lorentz gauge $\nabla^a(h_{ab}-1/2\eta_{ab}h^a_a)=0$ is
imposed, we get the following conditions on the metric
perturbations: $\Box v^b = 0$ and $\nabla_b \phi = 0$ (which
implies $\phi = \mathrm{constant}$). Any vector field
characterizing the remaining gauge transformations compatible
with the Lorentz gauge satisfies the condition $\Box \zeta^a=0$,
which implies $\Box V^a=0$ and $\nabla^a \Box \zeta=0$. We can
see that a vectorial gauge transformation compatible with the
Lorentz gauge can still be used to eliminate the vectorial part
(now both $v^a$ and $V^a$ must be solutions of the D'Alambertian
equation). On the other hand, a scalar gauge transformation such
that $\Box \zeta = - \phi = \mathrm{constant}$ (this is always
possible for Minkowski with a trivial --simply connected--
topology) can be introduced to get $\phi = 0$. Moreover, an
additional scalar gauge transformation compatible with the
Lorentz gauge and leaving $\phi$ invariant, which is
characterized by a $\zeta$ which satisfies the D'Alambertian
equation $\Box \zeta = 0$ (or, equivalently, $\tilde{\zeta}(p)=0$
for $p^2 \equiv p^\mu p^\nu \eta_{\mu\nu} \neq 0$ in Fourier
space), can be used to eliminate those contributions to $\psi$
which correspond to Fourier modes $\tilde{\psi}(p)$ with $p^2=0$
while leaving the remaining contributions unmodified. From now on
we will assume that the Lorentz gauge has been imposed and that
the additional gauge transformations just mentioned have been
carried out so that we are left only with the tensorial
components as well as those modes of the scalar component $\psi$
with $p^2 \neq 0$ in Fourier space.

One could select the gauge mentioned in the previous paragraph
imposing suitable conditions on the reduced Wigner functional
characterizing the initial state for the metric perturbations; see
Appendix~\ref{appC} for some additional comments on this point.
However, as explained in Appendix~\ref{appB}, asymptotic initial
conditions should be considered in order to get a finite result for
the metric correlation functions. Therefore, rather than fixing the
gauge for some initial state at some finite initial time, we will work
in Fourier space implicitly assuming asymptotic initial conditions and
fixing the gauge as described above.

In order to analyze the two-point quantum correlation function for
the metric perturbations, we will make use of the results mentioned in
Sec.~\ref{sec3} and described in some more detail
in Appendix~\ref{appE}. In particular, we will exploit the fact that the
stochastic correlation functions obtained with the solutions of the
Einstein-Langevin equation coincide with the quantum correlation
functions for the metric perturbations. Moreover, according to
Eq.~(\ref{correlation2}), the symmetrized two-point quantum
correlation function has two different contributions: the intrinsic
and the induced fluctuations. We proceed now to analyze each
contribution separately.

\subsubsection{Intrinsic fluctuations}
\label{sec6.1}

The first term on the right-hand side of Eq.~(\ref{correlation2})
corresponds to the fluctuations of the metric perturbations due to
the fluctuations of their initial state and is given by
\begin{equation}
\left\langle \Sigma_{ab}^{(0)}(x) \Sigma_{cd}^{(0)}(x') \right\rangle
_{\Sigma_{ab}^{(i)}, \Pi^{cd}_{(i)}}
\label{intrinsic},
\end{equation}
where we recall that $\Sigma_{ab}^{(0)}(x)$ is a solution of
the homogeneous part of the Einstein-Langevin equation (once the
Lorentz gauge has been imposed) with the appropriate initial
conditions.

As mentioned in Sec.~\ref{sec3} and Appendix~\ref{appE}, the
homogeneous part of the Einstein-Langevin equation actually
coincides with the linearized semiclassical Einstein
equation~(\ref{einstein3}). Therefore, we can make use of the results
derived in Refs.~\cite{horowitz80,flanagan96,anderson03}, which are
briefly summarized in Appendix~\ref{appC}. As described there, in
addition to the solutions with $G_{ab}^{(1)}(x)=0$, there are other
solutions that in Fourier space take the
form $\tilde{G}^{(1)}_{\mu\nu}(p) \propto \delta(p^2-p_0^2)$
for some particular values of $p_0^2$, but they all exhibit
exponential instabilities with Planckian characteristic timescales.

In order to deal with those unstable solutions, one possibility
is to employ the order reduction prescription. We are then left
only with the solutions which satisfy $\tilde{G}_{\mu\nu}^{(1)}(p)=0$
(see Appendix~\ref{appC}) . The result for the metric
perturbations in the gauge introduced above can be obtained by
solving for the Einstein tensor in that gauge:
$\tilde{G}^{(1)}_{ab}(p) = (1/2) p^2 (\tilde{h}_{\mu\nu} (p) -
1/2 \eta_{\mu\nu} \tilde{h}^\rho_\rho (p))$. Those solutions for
$\tilde{h}_{\mu\nu}(p)$ simply correspond to free linear
gravitational waves propagating in Minkowski spacetime expressed
in the transverse and traceless (TT) gauge. When substituting back
into Eq.~(\ref{intrinsic}) and averaging over the initial
conditions we simply get the symmetrized quantum correlation
function for free gravitons in the TT gauge for the state given
by the reduced Wigner function. As far as the intrinsic
fluctuations are concerned, it is clear that the order reduction
prescription is too drastic, at least in the case of Minkowski
spacetime, since no effects due to the interaction with the
quantum matter fields are left. The method employed in
Ref.~\cite{anderson03}, although slightly different, yields the
same result.

A second possibility, proposed by Hawking \emph{et al.}
\cite{hawking01,hawking02}, is to impose boundary conditions which
discard the runaway solutions that grow unboundedly in time and
correspond to a special prescription for the integration contour
when Fourier transforming back to spacetime coordinates (see
Appendix~\ref{appC} for a more detailed discussion). Following
that procedure we get, for example, that for a massless
conformally coupled scalar field with $\bar{\alpha} = 0$
\footnote{For the massless case one can always have $\bar{\alpha}
= 0$ by choosing the appropriate value of the renormalization
scale, as explained in Appendix~\ref{appD}.} and $\bar{\beta} > 0$
the intrinsic contribution to the symmetrized quantum correlation
function coincides with that of free gravitons plus an extra
contribution for the scalar part of the metric perturbations
$\phi$ which renders Minkowski spacetime stable but plays a
crucial role in providing a graceful exit for inflationary models
driven by the vacuum polarization of a large number of conformal
fields (such a massive scalar field would not be in conflict with
present observations because, for the range of parameters usually
considered, the mass would be far too large to have observational
consequences \cite{hawking01}).

\subsubsection{Induced fluctuations}
\label{sec6.2}

The second term on the right-hand side of Eq.~(\ref{correlation2})
corresponds to the fluctuations of the metric perturbations induced
by the fluctuations of the quantum matter fields and is given by
\begin{equation}
\frac{\bar{\kappa}^2}{N} \left( G_\mathrm{ret} \cdot \mathcal{N}
\cdot (G_\mathrm{ret})^{T} \right)_{abcd} (x,x')
= N \kappa^2 \left( G_\mathrm{ret} \cdot \mathcal{N}
\cdot (G_\mathrm{ret})^{T} \right)_{abcd} (x,x')
\label{induced},
\end{equation}
where $\mathcal{N}_{abcd}(x,x')$ is the noise kernel accounting for
the fluctuations of the stress tensor operator, and
$(G_\mathrm{ret})_{abcd}(x,x')$ is the retarded propagator with
vanishing initial conditions associated with the integro-differential
operator $L_{abcd}(x,x')$ defined in Eq.~(\ref{linearop}) of
Appendix~\ref{appE}.

As shown in Appendix~\ref{appE}, the symmetrized two-point quantum
correlation function coincides with the stochastic correlation
function obtained from the solutions of the Einstein-Langevin
equation. In fact, the contribution corresponding to the induced
quantum fluctuations, given by Eq.~(\ref{induced}), is equivalent to
the stochastic correlation function obtained by considering just the
inhomogeneous part of the solution to the Einstein-Langevin equation:
the second term on the right-hand side of Eq.~(\ref{solution2}).
Taking all that into account, it is clear that we can make use of the
results for the metric correlations obtained in Ref.~\cite{martin00}
by solving the Einstein-Langevin equation (the homogeneous part of the
solution was not considered there). In fact, one should simply take
$N=1$ to transform our expressions to those of Ref.~\cite{martin00}
and, similarly, multiply the noise kernel in the expressions of that
Reference by $N$ so that they can be used here, which follows
straightforwardly from the fact that we have $N$ independent matter
fields.

The same kind of exponential instabilities in the runaway
solutions of the homogeneous part of the Einstein-Langevin
equation (the linearized semiclassical Einstein equation) also
arise when computing the retarded propagator $G_\mathrm{ret}$. In
order to deal with those instabilities,  similar to the case of
the intrinsic fluctuations, one possibility is to make use of the
order reduction prescription. The Einstein-Langevin equation
becomes then $G_{ab}^{(1)} = \kappa \xi_{ab}$. The second
possibility, following the proposal of Hawking \emph{et al.}, is
to impose boundary conditions which discard the exponentially
growing solutions and translate into a special choice of the
integration contour when Fourier transforming back to spacetime
coordinates the expression for the propagator. In fact, it turns
out that the propagator which results from adopting that
prescription coincides with the propagator that was employed in
Ref.~\cite{martin00}. However, it should be emphasized that this
propagator is no longer the retarded one since it exhibits
causality violations at Planckian scales. A more detailed
discussion on all these points can be found in
Appendix~\ref{appC}.

Following Ref.~\cite{martin00}, the Einstein-Langevin equation can be
entirely written in terms of the linearized Einstein tensor
$\tilde{G}^{(1)}_{\mu\nu}(p)$ as follows:
\begin{equation}
F_{\mu\nu\alpha\beta}(p) \, \tilde{G}^{(1) \, \alpha\beta}(p)
= \bar{\kappa} \tilde{\xi}_{\mu\nu} (p) ,
\end{equation}
which simply corresponds to adding the stochastic source to the
linearized semiclassical Einstein equation (\ref{einstein3}),
where $F_{\mu\nu\alpha\beta}(p)$ was given by
Eq.~(\ref{Fkernel}). One can then solve the stochastic equation
for $\tilde{G}^{(1)}_{\mu\nu}(p)$ and obtain its correlation
function \cite{martin00}:
\begin{eqnarray}
\langle \tilde{G}^{(1)}_{\mu\nu}(p) \tilde{G}^{(1)}_{\rho\sigma}(p')
\rangle_\xi &=& \bar{\kappa}^2 \tilde{D}_{\mu\nu\alpha\beta}(p)
\langle \tilde{\xi}^{\alpha\beta}(p) \tilde{\xi}^{\gamma\delta}(p')
\rangle_\xi \tilde{D}_{\rho\sigma\gamma\delta}(p') \nonumber \\
&=& \frac{\bar{\kappa}^2}{N} \tilde{D}_{\mu\nu\alpha\beta}(p)
\tilde{\mathcal{N}}^{\alpha\beta\gamma\delta}(p)
\tilde{D}_{\rho\sigma\gamma\delta}(-p) (2\pi)^4 \delta(p+p')
\label{fourierGG1},
\end{eqnarray}
In the last equality  we have taken into account translational
invariance.  The noise kernel
$\tilde{\mathcal{N}}^{\alpha\beta\gamma\delta}(p)$ is given by
Eq.~(\ref{noiseMink}) in Appendix~\ref{appD}, and
$\tilde{D}_{\mu\nu\alpha\beta}(p)$ is the propagator that results from
inverting $F_{\mu\nu\alpha\beta}(p)$ (see Appendix~\ref{appC} for
a discussion on the uniqueness of this propagator) and is given by
\begin{equation}
\tilde{D}_{\mu\nu\alpha\beta}(p)
= \frac{1}{F_1(p)} \eta_{\mu(\alpha} \eta_{\beta)\nu}
- \frac{F_2(p)}{F_1(p) F_3(p)} p^2 P_{\mu\nu} \eta_{\alpha\beta} ,
\end{equation}
with $P_{\mu\nu} = \eta_{\mu\nu} - p_\mu p_\nu / p^2$, $F_1(p)$ and
$F_2(p)$ given by Eqs.~(\ref{F1}) and (\ref{F2}), and $F_3(p) = F_1(p)
+ 3 p^2 F_2(p)$.  On the other hand, if we make use of the order
reduction prescription, we get
\begin{equation}
\langle \tilde{G}^{(1)}_{\mu\nu}(p) \tilde{G}^{(1)}_{\rho\sigma}(p')
\rangle_\xi = \bar{\kappa}^2 \langle \tilde{\xi}^{\alpha\beta}(p)
\tilde{\xi}^{\gamma\delta}(p') \rangle_\xi = \frac{\bar{\kappa}^2}{N}
\tilde{\mathcal{N}}^{\alpha\beta\gamma\delta}(p) (2\pi)^4 \delta(p+p')
\label{fourierGG2}.
\end{equation}
Note that $G^{(1)}_{\mu\nu}(p)$ is gauge invariant when
perturbing a Minkowski background because the background tensor
$G^{(0)}_{ab}$ vanishes and, hence, $\mathcal{L}_{\vec{\zeta}}
G^{(0)}_{ab}$ also vanishes for any vector field $\vec{\zeta}$.

Finally, using the expression for the linearized Einstein tensor in
the Lorentz gauge, $\tilde{G}^{(1)}_{\mu\nu} = (1/2) p^2
\tilde{\bar{h}}_{\mu\nu}$ with $\bar{h}_{\mu\nu} = h_{\mu\nu}
- (1/2) \eta_{\mu\nu} h^{\alpha}_{\alpha}$, we obtain the correlation
function for the metric perturbations in that gauge:
\begin{equation}
\langle \tilde{\bar{h}}_{\mu\nu}(p) \tilde{\bar{h}}_{\rho\sigma}(p')
\rangle_\xi = \frac{4\bar{\kappa}^2}{N} \frac{1}{(p^2)^2}
\tilde{D}_{\mu\nu\alpha\beta}(p)
\tilde{\mathcal{N}}^{\alpha\beta\gamma\delta}(p)
\tilde{D}_{\rho\sigma\gamma\delta}(-p) (2\pi)^4 \delta(p+p')
\label{fourierhh1},
\end{equation}
or
\begin{equation}
\langle \tilde{\bar{h}}_{\mu\nu}(p) \tilde{\bar{h}}_{\rho\sigma}(p')
\rangle_\xi = \frac{4\bar{\kappa}^2}{N} \frac{1}{(p^2)^2}
\tilde{\mathcal{N}}_{\mu\nu\rho\sigma}(p) (2\pi)^4 \delta(p+p')
\label{fourierhh2},
\end{equation}
if the order reduction prescription is employed. It should be
emphasized that, contrary to the linearized Einstein tensor
$G^{(1)}_{ab}$, the metric perturbation $h_{ab}$ is not gauge
invariant. This should not pose a major problem provided that the
gauge has been completely fixed.

The correlation functions in spacetime coordinates can be easily
derived by Fourier transforming Eqs.~(\ref{fourierhh1}) or
(\ref{fourierhh2}). However, there is apparently an infrared
divergence at $p^2 = 0$, at least for the massless case. For the
massive case the result is finite because the noise kernel
$\tilde{\mathcal{N}}^{\alpha\beta\gamma\delta}(p)$ is
proportional to $\theta (-p^2 - 4m^2)$, so that $m^2 > 0$
guarantees that $p^2 = 0$ lies outside the domain of integration.
On the other hand, in the massless case the terms of the form
$p_\mu p_\nu p_\rho p_\sigma / (p^2)^2$ appearing when
substituting the noise kernel in Eqs.~(\ref{fourierhh1}) and
(\ref{fourierhh2}), give rise to infrared divergences when
computing the Fourier transform. In fact, even if we exclude the
massless case, the result would be finite, but it would become
larger and larger as we chose a positive but arbitrarily small
mass. In any case, such an infrared divergence seems to be
just a gauge artifact
\footnote{This is suggested by the fact that neither
the correlation function of the linearized Einstein tensor nor that
of the linearized Riemann tensor exhibits those divergences. The
finite result for the correlation function of the Einstein tensor
follows immediately from Eqs.~(\ref{fourierGG1}) and
(\ref{fourierGG2}), whereas for the Riemann tensor the potentially
divergent contributions coming from the terms proportional to
$p_\mu p_\nu p_\rho p_\sigma / (p^2)^2$ in the correlation
function for the metric perturbations involve exterior products
with $p_\alpha$ and, thus, vanish (of course the finite result
for the Einstein tensor could also have been inferred from the
finite result for the Riemann tensor). Alternatively, one can
eliminate the terms giving rise to divergences in
Eqs.~(\ref{fourierhh1}) and (\ref{fourierhh2}) by performing a
gauge transformation of the form $\tilde{h}_{\mu\nu}(p)
\rightarrow \tilde{h}_{\mu\nu}(p) + p_\mu p_\nu / p^2$, which is
generated by a vector field $\tilde{\zeta}^\mu (p) = p^\mu / p^2$
(consisting of just a scalar part $\tilde{\zeta} (p) = 1/p^2$).
Such a gauge transformation does not preserve the Lorentz
condition. Therefore, it seems that the infrared divergence is
simply indicating a singular massless limit for the Lorentz gauge
in the case under consideration.}.

We can conclude that, once the instabilities giving rise to the
unphysical runaway solutions have been properly dealt with, the
fluctuations of the metric perturbations around the Minkowski
spacetime induced by the interaction with quantum scalar fields
are indeed stable (if instabilities had been present, they would
have led to a divergent result when Fourier transforming back
to spacetime coordinates).
It should be emphasized that no ultraviolet
divergences related to the coincidence limit of the noise kernel
appeared in the previous analysis because we implicitly assumed
asymptotic initial conditions when working in Fourier space, as
explained in Appendix~\ref{appB}. Furthermore, in contrast to the
intrinsic fluctuations, even when using the order reduction
prescription there is still a nontrivial contribution to the
induced fluctuations due to the quantum matter fields.

\section{Discussion}
\label{sec7}

In this paper we make the point that an analysis of the stability
of any solution of SCG with respect to small
quantum corrections should consider not only the evolution of the
expectation value of the metric perturbations around that
solution, but also their fluctuations, encoded in the quantum
correlation functions. Making use of a large $N$ expansion, where
$N$ is the number of matter fields, the symmetrized two-point
quantum correlation function for the metric perturbations can be
decomposed into two distinct parts: the intrinsic fluctuations
due to the fluctuations of the initial state of the metric
perturbations itself, and the fluctuations induced by their
interaction with the matter fields. The stability of the first
contribution turns out to be closely related to the stability of
linearized perturbations of the semiclassical Einstein equation,
whereas the second contribution is equivalent to the stochastic
correlation functions in  stochastic semiclassical gravity
obtained from solutions of the Einstein-Langevin equation.

As a specific example, we analyzed the two-point quantum
correlation function for the metric perturbations around the
Minkowski spacetime interacting with $N$ scalar fields initially
in the Minkowski vacuum state. Once the ultraviolet instabilities
(discussed in Appendix~\ref{appC}) which are ubiquitous in
SCG and are commonly regarded as unphysical,
have been properly dealt with by using the order reduction
prescription or the procedure proposed in
Refs.~\cite{hawking01,hawking02}, both the intrinsic and the
induced contributions to the quantum correlation function for the
metric perturbations are found to be stable. In fact, one gets an
infrared divergence for the massless case when computing the
inverse Fourier transform for the induced contribution to the
correlation function of the metric, but that seems to be purely a
gauge effect, as argued in footnote~[86].

The symmetrized quantum correlation function for the metric
perturbations obtained is in agreement with the real part of the
propagator obtained by Tomboulis in Ref.~\cite{tomboulis77} using
a large $N$ expansion \footnote{The imaginary part can be easily
obtained from the expectation value for the commutator of the metric
perturbations, which is given by $2i\kappa(G_\mathrm{ret} (x',x)
- G_\mathrm{ret} (x,x'))$, as briefly explained in
Appendix~\ref{appE}.} (he actually considered
fermionic rather than scalar fields, but that just amounts to a
change in one coefficient). Tomboulis used the \emph{in-out}
formalism rather than the CTP formalism employed in this paper.
Nevertheless, his propagator is equivalent to the time-ordered
CTP propagator when asymptotic initial conditions are considered
because in Minkowski spacetime there is no real particle creation
and the \emph{in} and \emph{out} vacua are equivalent (up to some
phase which is absorbed in the usual normalization of the
\emph{in-out} propagator). The use of a CTP formulation is,
however, crucial to obtaining  true correlation functions rather
than transition matrix elements in dynamical (nonstationary)
situations (such as in an expanding Robertson-Walker
background geometry), where the \emph{in-out} scattering matrix
might not even be well defined at all.

As we pointed out in the Introduction, Ford and collaborators have
stressed the importance of the metric fluctuations and
investigated some of their physical implications
\cite{ford82,kuo93,ford99,ford03,borgman03,ford97,yu99,yu00}.
They have considered both intrinsic
\cite{ford97,ford99,yu99,yu00} and induced fluctuations
\cite{ford82,kuo93,ford99,ford03,borgman03}, which they usually
refer to as \emph{active} and \emph{passive} fluctuations
respectively. However, they usually consider these two kinds of
fluctuations separately and have not provided a unified treatment
where both of them can be understood as different contributions
to the full quantum correlation function.  Moreover, they always
neglect the nonlocal term which encodes the averaged back reaction
on the metric perturbations due to the modified dynamics of the
matter fields generated by the metric perturbations themselves
\footnote{In those References dealing with stochastic gravity
this term is usually called the dissipation term by analogy with
QBM models.}. Their justification is by arguing that those terms
would be of higher order in a perturbative expansion. That is
indeed the case when considering a Minkowski background if the
order reduction prescription is employed, but it is not clear
whether it remains true under more general conditions. In fact,
as mentioned in Ref.~\cite{roura03a}, for the usual cosmological
inflationary models the contribution of the nonlocal terms can be
comparable or even larger than that of the remaining terms.
Finally, in order to deal with the singular coincidence limit of
the noise kernel, in Ref.~\cite{kuo93} Ford and collaborators
opted to subtract a number of terms including the fluctuations
for the Minkowski vacuum. Even when no such subtraction was
performed (because a method based on multiple integrations by
parts was used instead) \cite{ford99,wu01,wu02}, they usually
discard the fluctuations for the Minkowski vacuum. Therefore, the
information on the metric fluctuations around a Minkowski
background when the matter fields are in the vacuum state is
missing in their work.

We close this Section by recalling a couple of partially open
issues for which either a better understanding or a better
treatment would be desirable. The first issue is the singular
coincidence limit for the noise kernel. It seems clear that, when
properly treating the noise kernel as a distribution, a finite
result for the metric correlation function is obtained except for
some divergent boundary terms at the initial time. There is a
natural physical interpretation: the completely uncorrelated
initial state that was considered becomes pathological when the
number of modes of the environment is infinite. A simple way to
overcome this problem and obtain a finite result for the
correlation function is to switch on the interaction smoothly so
that the modes of the environment with arbitrarily high
frequencies become correlated with the system in a nonsingular
way. However, in order to preserve the conservation of the source
in the Einstein-Langevin equation, which guarantees the
integrability of the equation through the Bianchi identity, the
interaction has to be turned on adiabatically and asymptotically
past initial conditions are required. Therefore, other procedures
should be devised to address situations that require specifying
the initial conditions at a finite initial time.

The other question which deserves further study is the procedure
employed to deal with the runaway solutions discussed in
Appendix~\ref{appC}. The order reduction prescription is rather
drastic as its net outcome is to discard entirely the contribution
from the dissipation kernel (as far as the expectation value and
the symmetrized two-point correlation function are concerned),
which encodes the averaged back reaction of the matter fields on
the metric perturbations.
As for the method employed by Hawking \emph{et al.} in
Refs.~\cite{hawking01,hawking02}, we find the fact that the
choice of the physical solutions at a given instant of time
depends on the far future somewhat unsatisfactory, and
discarding solutions which grow unboundedly in time could get
rid of other possible instabilities which are physically meaningful.
Furthermore, it is not clear whether both procedures could be
implemented in a general case.

To gain insight into some of the previous aspects, an interesting
possibility is to consider an analogous situation in QED
with the electromagnetic field regarded as an open quantum system
interacting with an environment constituted by the charged quantum
fields. In fact, the analogy between SCG and
the equation for the expectation value of the electromagnetic
field to leading order (order 1) in a large $N$ expansion for $N$
charged quantum fields has been discussed by a number of
authors \cite{horowitz81,hartle81,jordan87a}. One step further
was considered in Ref.~\cite{cooper94}, where the evolution of the
expectation value of the electromagnetic field was considered to
next to leading order in $1/N$ (order $1/N$). The two-point
quantum correlation functions (the CTP propagators) for the
electromagnetic field to leading order in $1/N$ (order $1/N$),
which play a crucial role there, are completely analogous to the
quantum correlation functions for the metric perturbations
considered here.

\begin{acknowledgments}
It is a pleasure to thank Enrique \'{A}lvarez, Paul Anderson, Daniel
Arteaga, Dieter Brill, Larry Ford, Carmen Molina-Par\'{\i}s and Emil
Mottola for interesting discussions.
B.~L.~H. and A.~R. are supported by NSF under Grant PHY03-00710.
E.~V. acknowledges support from the MICYT Research Project
No.~FPA-2001-3598.
\end{acknowledgments}

\appendix

\section{Intrinsic and induced fluctuations in a simple quantum
Brownian motion model}
\label{appA}

In this Appendix we illustrate the importance of the fluctuations
induced by the environment when considering the quantum fluctuations
for an open system. As an example we will use a simple model which was
analyzed in some detail in Ref.~\cite{calzetta03a}: a linear quantum
Brownian motion (QBM) model that consists of a harmonic oscillator,
which will be referred to as the system, bilinearly coupled to a set
of harmonic oscillators, which constitute the environment.

In Ref.~\cite{calzetta03a} it was shown that a stochastic
description based on a Langevin type equation could be used to
gain information on the quantum properties of the open system. In
particular, the symmetrized two-point quantum correlation
function for the system turns out to be equivalent to the
correlation function obtained in the context of the stochastic
description:
\begin{equation}
\frac{1}{2}\left\langle\left\{\hat{x}(t_1),\hat{x}(t_2)\right\}\right\rangle
=\left\langle \left\langle X(t_{1})X(t_{2})\right\rangle _{\xi }\right\rangle
_{X_{i},p_{i}},
\label{correlQBM}
\end{equation}
where $\langle \ldots \rangle$ denotes the expectation value with
respect to the quantum state of the system, $\hat{x}(t)$ is the
position operator for the system in the Heisenberg picture,
$\langle \ldots \rangle_\xi$ denotes the average over all possible
realizations of the stochastic source $\xi(t)$ and $\langle \ldots
\rangle_{X_{i},p_{i}}$ is the average over all possible initial
conditions for the solutions of the Langevin equation distributed
according to the reduced Wigner function for the initial state of
the system. The functions $X(t)$ appearing inside the stochastic
averages are solutions of the Langevin equation $L \cdot X =
\xi$, where $L(t,t') = M (d^{2}/dt^2 + \Omega_\mathrm{ren}^2)
\delta(t-t') + H_\mathrm{ren}(t,t')$, with $H_\mathrm{ren}$ being
the renormalized kernel appearing in the real part of the
influence action and $\cdot \equiv \int_{t_i}^{t_f}dt$ throughout
this Appendix. Here $\xi(t)$ is a Gaussian stochastic source with
vanishing expectation value and correlation function $\langle
\xi(t)\xi(t') \rangle_\xi = N(t,t')$ where $N(t,t')$ is  the
noise kernel, being the kernel appearing in the imaginary part of
the influence action (see Ref.~\cite{calzetta03a} for further
details).  When the environment is initially in a thermal
equilibrium state, the noise kernel is explicitly given by
$N(t,t') = \int_0 ^\infty d\omega I(\omega) \coth \beta\omega
\cos \omega (t-t')$, where $I(\omega)$ is the spectral density
function, which characterizes the frequency distribution of the
oscillators in the environment.

The solution of the Langevin equation can be written as
$X(t)=X_0(t) + (G_\mathrm{ret} \cdot \xi)(t)$, where $X_0(t)$ is
a solution of the homogeneous part of the Langevin equation which
contains all the information about the initial conditions and
$G_\mathrm{ret}(t,t')$ is the retarded propagator with vanishing
initial conditions. Substituting the previous expression for
$X(t)$ into Eq.~(\ref{correlQBM}) and using the properties of the
stochastic source, one  obtains the following result for the
two-point quantum correlation function:
\begin{equation}
\frac{1}{2}\left\langle\left\{\hat{x}(t_1),\hat{x}(t_2)\right\}\right\rangle
=\left\langle X_0(t_1)X_0(t_2) \right\rangle_{X_{i},p_{i}} +
\left(G_\mathrm{ret} \cdot N \cdot \left(G_\mathrm{ret}\right)^{T}\right)
(t_1,t_2) \label{correlQBM2},
\end{equation}
where the first contribution corresponds to the \emph{intrinsic}
fluctuations connected to the dispersion of the initial state of
the system, and the term proportional to the noise kernel reflects
the fluctuations \emph{induced} by the system's interaction with
the environment. Note the close analogy between
Eq.~(\ref{correlQBM2}) and the expression for the symmetrized
two-point quantum correlation function in the gravitational case,
given by Eq.~(\ref{correlation2}).

Let us specialize to the case of an ohmic environment,
\emph{i.e.}, the case in which the spectral distribution function
for the frequencies of the oscillators in the environment is of
the form $I(\omega)=M \gamma \omega$, where $M$ is the mass of
the system harmonic oscillator and $\gamma$ is some constant
proportional to the square of the system-environment coupling
constant. Then the kernel $H_\mathrm{ren}$ becomes local with
$H_\mathrm{ren}(t,t') = M \gamma \delta'(t-t')$ and the
homogeneous solution $X_0(t)$ takes on the following simple form:
\begin{equation}
X_0(t) = e^{-\frac{\gamma}{2}(t-t_i)} \left[X_i \cos\tilde{\Omega}(t-t_i)
+ \left(\frac{p_i}{M \tilde{\Omega}} + \frac{\gamma}{2\tilde{\Omega}} X_i\right)
\sin\tilde{\Omega}(t-t_i) \right],
\end{equation}
where $\tilde{\Omega}=\sqrt{\Omega^2_\mathrm{ren}-(\gamma/2)^2}$
and we considered the underdamped case ($\Omega_\mathrm{ren} >
\gamma/2$). A similar result also holds for the overdamped case
with the trigonometric functions replaced by the hyperbolic
functions. Due to the exponential factor, $X_0(t)$ and hence the
intrinsic fluctuations will decay at times much larger than the
relaxation time $2 \gamma^{-1}$ \footnote{The existence of such a
decay still holds for the overdamped case provided that
$\Omega_\mathrm{ren} \neq 0$, otherwise there is a constant
contribution that does not decay in time.}. In fact, if we take
the limit $t_i \rightarrow -\infty$, the contribution to the
two-point correlation function from the intrinsic fluctuations
completely vanishes and one is just left with the induced
fluctuations. If the initial state of the environment was a
thermal state, the dissipation kernel (the antisymmetric part of
$H_\mathrm{ren}(t,t')$) and the noise kernel are
related by a fluctuation-dissipation relation which characterizes
the balance between the noise induced by the environment and
the dissipation effect so that the two-point correlation function
remains bounded in time.

{F}rom the example employed in this Appendix, it is clear that the
induced fluctuations play an important role when considering correlation
functions in open quantum systems. In fact, for asymptotically past
initial conditions they become the entire contribution to the
correlation function since the intrinsic fluctuations are completely
damped by the dissipation. In a more general context, such as the
gravitational case, the dissipation kernel will not damp the intrinsic
fluctuations, but the induced ones will still play an important role.

\section{Dissipation and noise kernels in Minkowski spacetime}
\label{appD}

In this Appendix we provide the expressions for the dissipation and
noise kernels of a free real scalar field when a Minkowski background
spacetime is considered and the state of the fields is the Minkowski
vacuum. The details of their derivation can be found in
Ref.~\cite{martin00}. All the expressions in this Appendix are given
in Fourier space and derived by making use of the translational invariance
in terms of the inertial coordinates employed for the Minkowski
background. Given any expression $\tilde{A}(p)$, the corresponding
expression in spacetime coordinates $A(x-y)$ can be simply obtained from
\begin{equation}
A(x-y) = \int \frac{d^4 p} {(2\pi)^4} \, e^{i px} \, \tilde{A}(p) .
\end{equation}

The linear combination $-2(M+H)$ with $M$ and $H$ given by
Eqs.~(\ref{Hkernel})-(\ref{Mkernel}) is commonly referred to as the
polarization tensor in analysis based on linear response theory. We
will use the term ``dissipation kernel'', by analogy with the usual
terminology employed in the context of open quantum systems
\footnote{Strictly speaking, the term dissipation kernel commonly
refers to the antisymmetric part of the kernel $-H_\mathrm{ren}$.
Making an abuse of language, we will employ this term to refer to
the whole kernel $-2H_\mathrm{ren}$ plus the local and symmetric
kernel $-2M_\mathrm{ren}$.}.
In the case under consideration the local kernel
$M^{\mu\nu\alpha\beta}_\mathrm{ren}(x-y)$ is proportional to the
Einstein tensor and can be absorbed in a finite renormalization of
the gravitational coupling constant. The expression
in Fourier space for the nonlocal kernel
$H^{\mu\nu\alpha\beta}_\mathrm{ren}(x-y)$ is
\begin{equation}
\tilde{H}^{\mu\nu\alpha\beta}_\mathrm{ren}(p) = \frac{1}{2}
\left( P^{\mu\alpha} P^{\nu\beta} - \frac{1}{3}
P^{\mu\nu}P^{\alpha\beta} \right)
\left(\tilde{H}_A(p) - 2\bar{\alpha} \right) + P^{\mu\nu}P^{\alpha\beta}
\left( \tilde{H}_B(p) - 2\bar{\beta} \right)
\label{Hkernel2},
\end{equation}
where $\bar{\alpha}$ and $\bar{\beta}$ are constants which include
the renormalized parameters $\alpha$ and $\beta$ appearing in
Eq.~(\ref{einstein1}), $P_{\mu\nu}$ is the projector orthogonal to
$p^\mu$, given by $P_{\mu\nu} = \eta_{\mu\nu} - p_\mu p_\nu / p^2$,
and
\begin{eqnarray}
\tilde{H}_A(p) = \frac{1}{1920 \pi^2} \Biggl\{\left(1 + 4
\frac{m^2}{p^2} \right)^2 \Biggl[ - i \pi \mathrm{sign} \, p^0 \;
\theta (-p^2 - 4 m^2) \, \sqrt{1 + 4 \frac{m^2}{p^2}}
+ \varphi(p^2) \Biggr] - \frac{8}{3} \frac{m^2}{p^2} \Biggr\}
\label{HAkernel}, \\
\tilde{H}_B(p) = \frac{1}{288 \pi^2} \Biggl\{
\left(3 \left( \xi - \frac{1}{6} \right)
+ \frac{m^2}{p^2} \right)^2 \Biggl[ - i \pi \mathrm{sign} \, p^0 \;
\theta (-p^2 - 4m^2) \, \sqrt{1 + 4 \frac{m^2}{p^2}} + \varphi(p^2)
\Biggr] - \frac{1}{6} \frac{m^2}{p^2} \Biggr\}
\label{HBkernel},
\end{eqnarray}
where $\xi$ is the parameter characterizing the coupling of the scalar
field to the spacetime curvature through a term of the form $- (\xi /
2) R \phi^2 $ in the matter Lagrangian, and $\varphi(p^2)$ is given by
\begin{eqnarray}
\varphi (p^2) =  \int_0^1 d\alpha \, \ln \biggl| 1
+ \frac{p^2}{m^2} \alpha (1 - \alpha) \biggr|
&=& -2 + \sqrt{1 + 4 \frac{m^2}{p^2}} \, \ln \left|
\frac{\sqrt{1 + 4 \frac{m^2}{p^2}} + 1}
{\sqrt{1 + 4 \frac{m^2}{p^2}} - 1} \right|
\theta \biggl(1 + 4 \frac{m^2}{p^2} \biggr) \nonumber \\
&& + 2 \sqrt{-1 - 4 \frac{m^2}{p^2}} \,
\mathrm{arccotan} \Biggl( \sqrt{-1 - 4 \frac{m^2}{p^2}} \Biggr) \,
\theta \biggl(-1 - 4 \frac{m^2}{p^2} \biggr)
\label{varphi}.
\end{eqnarray}
Using the renormalized version of Eq.~(\ref{STexpecvalue}) in Fourier
space, the dependence on the metric of the renormalized expectation
value of the stress tensor operator can be written entirely in terms
of the linearized Einstein tensor as follows:
\begin{equation}
\left\langle \widetilde{\hat{T}^{\mu\nu}}[g+h;p) \right\rangle_\mathrm{ren}
= 2 P^{\mu\nu} \left( -\frac{1}{3} \tilde{H}_A(p) + \frac{2}{3}
\bar{\alpha} - \tilde{H}_B(p) + 2\bar{\beta} \right)
\left(\tilde{G}^{(1)}\right)^\alpha_\alpha (p) + \frac{2}{3}
P^\alpha_\alpha \left( \tilde{H}_A(p) - 2\bar{\alpha} \right)
\left(\tilde{G}^{(1)}\right)^{\mu\nu} \label{STexpecvalue2}.
\end{equation}
Following Ref.~\cite{martin00}, we have employed a renormalization
scheme in which the renormalization scale is fixed to $\mu^2 =
m^2$. This is, of course, not possible for the massless
case. Nevertheless, the expression for the massless case can still be
obtained by adding a term $\ln (m^2 / \mu^2)$ to Eq.~(\ref{varphi})
and subtracting $(1920 \pi^2)^{-1} \ln (\mu^2 / m^2)$ and $(\xi -
1/6)^2  (96 \pi^2)^{-1} \ln (\mu^2 / m^2)$ respectively from $2
\bar{\alpha}$ and $2 \bar{\beta}$ in Eqs.~(\ref{Hkernel2}) and
(\ref{STexpecvalue2}) before taking the limit $m^2 \rightarrow 0$. The
renormalized parameters will then depend on the arbitrary scale
$\mu$. If desired, it is always possible to choose $\bar{\alpha} = 0$
by fixing the renormalization scale $\mu$ to some appropriate value.

Finally, the expression for the noise kernel in Fourier space is
given by
\begin{eqnarray}
\tilde{\mathcal{N}}_{\mu\nu\rho\sigma}(p)
&=& \frac{1}{2880 \pi} \theta (-p^2 - 4m^2) \, \sqrt{1 + \frac{4 m^2}{p^2} }
\left[ \frac{1}{4} \left(1 + \frac{4 m^2}{p^2} \right)^2 (p^2)^2 \,
\bigl(3 P_{\mu (\rho}P_{\sigma )\nu} - P_{\mu\nu} P_{\rho\sigma} \bigr)
\right. \nonumber \\
&& \left. + 10 \left(3 \left( \xi - \frac{1}{6} \right)
+ \frac{m^2}{p^2} \right)^2
(p^2)^2 P_{\mu\nu} P_{\rho\sigma} \right]
\label{noiseMink} .
\end{eqnarray}

\section{Stochastic and quantum correlation functions}
\label{appE}

It was initially believed that some kind of environment-induced
decoherence mechanism was required to realize the stochastic dynamics
described by the Einstein-Langevin equation
\cite{martin99c,martin99b}. Later, in Ref.~\cite{calzetta03a} it
was shown that, even in the absence of decoherence, a stochastic
description based on a Langevin type equation contains nontrivial
information on fully quantum properties of simple linear open
quantum systems. In particular, the reduced Wigner function of the
system (see, for instance, Ref.~\cite{hillary84} for the definition
and properties of the Wigner function) can be expressed as a double
average for the solutions of the Langevin equation with respect to
both the different realizations of the stochastic source and the
initial conditions, which are distributed according to the reduced
Wigner function at the initial time. This expression can then be used
to derive the master equation governing the time evolution of the
reduced Wigner function (or, equivalently, the reduced density
matrix). Furthermore, the stochastic correlation functions for the
solutions of the Langevin equation are actually equivalent to quantum
correlation functions for the system observables.

Although the previous results were obtained in Ref.~\cite{calzetta03a}
for linear open quantum systems, they can be extended to the case of
nonlinear quantum field theories provided that some kind of Gaussian
approximation for the corresponding influence functional is
considered. In fact, in Ref.~\cite{roura03b} it will be explained in
detail how those results can indeed be shown to hold for the metric
fluctuations around a given background spacetime by properly treating
the gauge freedom and the corresponding dynamical
constraints. More precisely, when considering $N$ free quantum matter
fields weakly interacting with the gravitational field in the sense that the
gravitational coupling constant times the number of fields remains
constant in the large $N$ limit, the stochastic correlation functions
can be shown to coincide with the leading order contribution to the
quantum correlation functions of the metric perturbations in a large $N$
expansion.

Here we briefly sketch, in the context of a large $N$ expansion, some
of the key aspects in the derivation of the result stated above. The
details will appear in Ref.~\cite{roura03b} and were partially
included in Ref.~\cite{roura01}.

We will consider metric perturbations around a globally hyperbolic
background spacetime (that will be specialized to Minkowski
spacetime at some point) regarded as an open quantum system
interacting with the quantum matter fields, which constitute the
environment. In particular we will consider $N$ minimally coupled
free scalar fields, but the main result can be generalized to
nonminimally coupled scalar fields or even vectorial and
fermionic fields. The action for the combined system is the sum
of the gravitational action $S_\mathrm{g}$ plus the action for the
matter fields $S_\mathrm{m}$. The gravitational action is given
by the usual Einstein-Hilbert term, the corresponding boundary
term (which should be included to have a well-defined variational
problem and will later be important) and the usual counterterms
required to renormalize the divergences arising when functionally
integrating the matter fields:
\begin{equation}
S_\mathrm{g} = \frac{N}{2\bar{\kappa}} \int_\mathcal{M} d^4x \sqrt{-\tilde{g}}
R\left(\tilde{g}\right) +\frac{N}{\bar{\kappa}}
\int_{\mathcal{S}=\partial\mathcal{M}}d^3x
\sqrt{\tilde{g}_{\mathcal{S}}}K_a^a(\tilde{g}) +\
(\mathrm{counterterms}) ,
\end{equation}
where $\tilde{g}_{ab}=g_{ab}+h_{ab}$ is the perturbed metric, $g_{ab}$
is the background metric and the gravitational coupling constant
$\kappa = 8\pi / m_p^2$ was rescaled to $\bar{\kappa}/N$ so that the
product of the rescaled gravitational constant times the number of
fields remains constant in the limit $N \rightarrow \infty$. The
action for the matter fields is
\begin{equation}
S_\mathrm{m} = -\sum_{j=1}^N \int_\mathcal{M} d^4x\sqrt{-\tilde{g}}\frac{1}{2}
\left( \tilde{g}^{ab} \nabla_a\varphi_j \nabla_b\varphi_j
+ m^2 \varphi_j^2 \right) ,
\end{equation}
where $m$ is the mass of the scalar field. In fact, we will not take
the limit $N \rightarrow \infty$, but rather use the expansion in
$1/N$ as a useful way to organize our computation and the
contributions that are included. At the very end one can always
substitute back the rescaled gravitational constant in terms of the
physical one.

The closed time path (CTP) generating functional for the metric
perturbations, from which true expectation values and correlation
functions can be obtained
\cite{schwinger61,keldysh65,chou85,jordan86,calzetta87,campos94},
is given by
\begin{eqnarray}
Z_\mathrm{CTP}[J_{ab},J'_{cd}] &=& \int \mathcal{D}h_{ab} \mathcal{D}h'_{cd}
e^{i S_\mathrm{g}[h]-i S_\mathrm{g}[h']} e^{i J \cdot h - i J' \cdot h'}
\rho_\mathrm{r}[h_{ab}^{(i)},h_{cd}^{\prime\,(i)}] \nonumber \\
&& \times \prod_{j=1}^{N} \int \mathcal{D}\varphi_j \mathcal{D}\varphi'_j
e^{i S_\mathrm{m}[\varphi_j,h] -i S_\mathrm{m}[\varphi'_j,h']}
\rho[\varphi_j^{(i)},\varphi_j^{\prime\,(i)}] ,
\end{eqnarray}
where we have used the notation $A\cdot B\equiv \int d^{4}y
\sqrt{-g(y)} A^{ab}(y) B_{ab}(y)$. The density matrices for the
initial state of the fields and the metric perturbations, which are
all assumed to be initially uncorrelated, are
$\rho[\varphi_j^{(i)},\varphi_j^{\prime\,(i)}]$ and
$\rho[h_{ab}^{(i)},h_{cd}^{\prime\,(i)}]$ respectively. The gauge
freedom in the path integrals for the metric perturbations should be
properly treated, as briefly described below.

The first step is to integrate out the matter fields using the
influence functional formalism of Feynman and Vernon for open
quantum systems \cite{feynman63,feynman65}. The influence action
$S_\mathrm{IF}$ is defined as
\begin{equation}
e^{i S_\mathrm{IF}[h,h']} = \prod_{j=1}^{N} \int \mathcal{D}\varphi_j
\mathcal{D}\varphi'_j
e^{i S_\mathrm{m}[\varphi_j,h] -i S_\mathrm{m}[\varphi'_j,h']}
\rho[\varphi_j^{(i)},\varphi_j^{\prime\,(i)}] .
\end{equation}
Up to quadratic order in the metric perturbations it is given by
\cite{martin99b,martin99c}
\begin{equation}
S_\mathrm{IF}\left[\Sigma_{ab},\Delta_{ab}\right]
= N \left( Z\cdot \Delta + \Delta \cdot (H+M) \cdot \Sigma +
\frac{i}{8} \Delta \cdot \mathcal{N} \cdot \Delta \right) ,
\end{equation}
where we have introduced the semisum and difference variables
$\Sigma_{ab} = \left(h_{ab} + h'_{ab}\right)/2$ and
$\Delta_{ab}=h'_{ab} - h_{ab}$, $Z^{ab}(x) = -(1/2)
\langle\hat{T}^{ab}[\hat{\varphi},g;x) \rangle$ and the kernels $H$,
$M$ and $\mathcal{N}$ were defined in Eqs.~(\ref{Hkernel}),
(\ref{Mkernel}) and (\ref{noise}). As explained in Sec.~\ref{sec2},
the kernels $H$ and $M$ exhibit divergences that are canceled by
renormalizing the gravitational coupling constant and the cosmological
constant in the bare gravitational action as well as the coupling
constants of the counterterms quadratic in the curvature. We will not
need terms of higher order in the metric perturbations because they
give contributions to the connected part of the CTP generating
functional (given by $W^\mathrm{(LO)}_\mathrm{CTP} = -i \ln
Z^\mathrm{(LO)} _\mathrm{CTP}$) of higher order in $1/N$.  This is
also true for the terms in the gravitational action $S_\mathrm{g}$,
which implies that we do not have to consider graviton vertices.  In
order to show that, when computing the connected part of the CTP
generating functional to leading order in $1/N$, it is indeed
sufficient to keep just those terms in the gravitational action and
the influence action which are at most quadratic in the metric
perturbations, one can first compute the generating functional
resulting from that approximation, $Z_\mathrm{CTP}^\mathrm{(LO)}$, and
then show that including terms of higher order in the metric
perturbations would yield corrections to the connected part of the
generating functional of higher order in $1/N$.

We now specialize to the case of a Minkowski background,
\emph{i.e.}, $g_{ab}=\eta_{ab}$, and consider a family of Cauchy
hypersurfaces which foliate the spacetime into constant time
hypersurfaces of a given inertial frame in Minkowski spacetime.
The initial states for the metric perturbations and the matter
fields are specified on one of these hypersurfaces, which will be
denoted by $\mathcal{S}_i$. Another hypersurface is chosen as the
final hypersurface $\mathcal{S}_f$ so that any spacetime region of
interest lies between them. Then we integrate by parts those
contributions to the Einstein-Hilbert term of the gravitational
action involving two derivatives acting on the same factor and
impose the Lorentz gauge condition $\nabla^a \bar{h}_{ab}=0$
(we recall that the indices are raised and lowered using the
background Minkowski metric, all the covariant derivatives are
taken in this background metric and $\bar{A}_{ab} \equiv
A_{ab} - (1/2) \eta_{ab} A^a_a$). The boundary terms resulting
from integration by parts are canceled by the boundary terms
included in the gravitational action \footnote{In general, one
should be careful with the contributions from the timelike
boundaries  as well as the edges connecting the spacelike and
timelike boundaries \cite{pons03}. Here we will assume that the
timelike boundaries are infinitely far away and the value of the
metric perturbations decays at large distances so that only the
contributions from the spacelike boundaries are relevant.} and
the expression $\tilde{S}_\mathrm{g}[h,h'] =
S_\mathrm{g}[h]-S_\mathrm{g}[h']$ up to quadratic order in the
metric perturbations becomes
\begin{equation}
\tilde{S}_\mathrm{g}[\Sigma_{ab},\Delta_{cd}]
= \frac{N}{4\bar{\kappa}}\int_\mathcal{M}
d^{4}x \sqrt{-g} \nabla_a \Delta^{bc} \nabla^a \overline{\Sigma}_{bc}
+\ (\mathrm{counterterms})  \label{Sg2}.
\end{equation}
Next, we introduce the momentum canonically conjugate to
$\Delta_{ab}$ \footnote{Throughout this Section we will neglect the
contribution to the momentum from the counterterms; see
Appendix~\ref{appC} for further discussion on this point.}, which is given by
\begin{equation}
\Pi^{ab} [\Sigma_{cd}] = \frac{\delta \tilde{S}_\mathrm{g}}
{\delta \dot{\Delta}_{ab}} = - \frac{N}{4\bar{\kappa}}
\dot{\overline{\Sigma}}^{ab} ,
\end{equation}
where we employed the notation $\dot{A}_{ab} \equiv n^c \nabla_c
A_{ab}$ for the covariant derivative with respect to the normalized
and future-directed timelike vector $n^a$ orthogonal to the family of
Cauchy hypersurfaces including the initial and final hypersurfaces
$\mathcal{S}_i$ and $\mathcal{S}_f$. Finally, one can integrate again
by parts so that
\begin{eqnarray}
S_\mathrm{g}[\Sigma_{ab},\Delta_{cd}] &=& -\frac{N}{4\bar{\kappa}}
\int_\mathcal{M} d^{4}x \sqrt{-g}
\Delta^{bc} \nabla_a  \nabla^a \overline{\Sigma}_{bc}
+ \int_{\mathcal{S}_f=\partial\mathcal{M}} d^3x
\sqrt{g_{\mathcal{S}_f}} \Pi_{(f)}^{ab} \Delta_{ab}^{(f)}
- \int_{\mathcal{S}_i=\partial\mathcal{M}} d^3x
\sqrt{g_{\mathcal{S}_i}} \Pi_{(i)}^{ab} \Delta_{ab}^{(i)}
\nonumber \\
&& +\ (\mathrm{counterterms})
\label{Sg3} ,
\end{eqnarray}
where the indices $(i)$ and $(f)$ denote quantities evaluated on
$\mathcal{S}_i$ and $\mathcal{S}_f$ respectively. It should be
emphasized that $h_{ab}^{(i)}$ or $h_{ab}^{(f)}$ simply correspond to
the spacetime metric evaluated on those hypersurfaces and should not
be confused with the induced metric in the usual ADM
formulation. Furthermore, the contribution for $\mathcal{S}_f$ will
not be relevant because, when computing the CTP generating functional,
we should take $h_{ab}=h'_{ab}$ (which implies $\Delta_{ab}=0$) on the
final hypersurface.

Changing to the new current variables $J_{ab}^{\Sigma} = (J_{ab} +
J'_{ab})/2$ and $J_{ab}^{\Delta} = J'_{ab} - J_{ab}$, and functionally
integrating with respect to $\Delta_{cd}$, one gets the
following expression for the generating functional:
\begin{equation}
Z^\mathrm{(LO)}_\mathrm{CTP}[J_{ab}^{\Sigma},J_{cd}^{\Delta}]
= K \int \mathcal{D}\Sigma_{ab} e^{-\frac{N}{2 \bar{\kappa}^2}
\left(L \cdot \Sigma - \frac{2 \bar{\kappa}}{N}J^{\Sigma} \right)
\cdot \mathcal{N}^{-1} \cdot \left(L \cdot \Sigma
- \frac{2 \bar{\kappa}}{N}J^{\Sigma}\right)} e^{i J^{\Delta} \cdot \Sigma}
W_\mathrm{r}[\Sigma_{ab}^{(i)},\Pi^{cd}_{(i)}] \label{ZCTP3},
\end{equation}
where the functional integral with respect to $\Sigma_{ab}$ is
restricted to those configurations that satisfy the Lorentz gauge
condition, and we introduced the integro-differential operator
\begin{equation}
L_{abcd}(x,x') = (1/2) (\eta_{ac}\eta_{bd} - \eta_{ab}\eta_{cd}/2)
\Box \delta(x-x') + 2 \bar{\kappa} H^\mathrm{(ren)}_{abcd}(x-x')
+ 2 \bar{\kappa} M^\mathrm{(ren)}_{abcd}(x-x') \label{linearop}.
\end{equation}
$K$ is some normalization constant which can be eventually determined
by demanding that $Z^\mathrm{(LO)}_\mathrm{CTP} [J_{ab}^{\Sigma},
J_{cd}^{\Delta}] = 1$ when we take $J_{ab}^{\Sigma} = J_{cd}^{\Delta}
= 0$, and $W_\mathrm{r}[\Sigma_{ab}^{(i)},\Pi^{cd}_{(i)}]$ is the reduced
Wigner functional for the metric perturbations at the initial time,
which is defined in terms of the reduced density matrix at the initial
time as
\begin{equation}
W_\mathrm{r}[\Sigma_{ab}^{(i)},\Pi^{cd}_{(i)}] = (1/2\pi) \int
d\Delta_{cd}^{(i)} \exp(i \Pi^{cd}_{(i)} \Delta_{cd}^{(i)})
\rho_\mathrm{r}[\Sigma_{ab}^{(i)}, \Delta_{cd}^{(i)}] \label{wigner} .
\end{equation}
The Lorentz gauge condition does not fix completely the gauge
freedom under local diffeomorphisms. However, it can be completely
fixed by imposing additional gauge fixing conditions on the state of
the metric perturbations at the initial time. Thus, from now on it
should be understood that some appropriate condition such as the
transverse and traceless gauge has been imposed on the reduced
density matrix (or, equivalently, the reduced Wigner functional) at
the initial time.

Introducing a suitable functional change, the CTP generating
functional can be rewritten in the following way when taking
$J_{ab}^{\Sigma} = 0$:
\begin{equation}
Z^\mathrm{(LO)}_\mathrm{CTP}[J_{ab}^{\Sigma} = 0, J_{cd}^{\Delta}]
= \left\langle\left\langle  e^{i J^{\Delta} \cdot \Sigma} \right\rangle_\xi
\right\rangle_{\Sigma_{ab}^{(i)}, \Pi^{cd}_{(i)}} \label{ZCTP5},
\end{equation}
where the expectation values $\left\langle \ldots \right\rangle
_{\Sigma_{ab}^{(i)}, \Pi^{cd}_{(i)}}$ and $\left\langle \ldots
\right\rangle_\xi$ are defined as
\begin{eqnarray}
\left\langle \ldots \right\rangle_{\Sigma_{ab}^{(i)}, \Pi^{cd}_{(i)}}
= \int d\Sigma_{ab}^{(i)} d\Pi^{cd}_{(i)} \ldots
W_\mathrm{r}[\Sigma_{ab}^{(i)},\Pi^{cd}_{(i)}] \label{expectation1}, \\
\left\langle \ldots \right\rangle_\xi = (2\pi \mathcal{N}/N)^{-1/2}
\int \mathcal{D}\xi_{ab} \ldots e^{- \frac{N}{2} \xi \cdot
\mathcal{N}^{-1} \cdot \xi} \label{expectation2}.
\end{eqnarray}
The $\Sigma_{ab}(x)$ inside the expectation values in
Eq.~(\ref{ZCTP5}) satisfies the equation
\begin{equation}
(L \cdot \Sigma)_{ab} (x) = \bar{\kappa} \xi_{ab} (x) \label{langevin2},
\end{equation}
with initial conditions $\Sigma_{ab}^{(i)}$ and
$\dot{\Sigma}_{ab}^{(i)} = - (4\bar{\kappa} / N)
\overline{\Pi}^{cd}_{(i)}$ on the initial hypersurface
$\mathcal{S}_i$. From Eq.~(\ref{expectation2}) it becomes clear that
one can formally interpret $\xi_{ab}$ as a Gaussian stochastic source
with a vanishing expectation value and whose correlation function is
given by the noise kernel. Eq.~(\ref{langevin2}) can then be regarded
as a stochastic Langevin equation and coincides with the
Einstein-Langevin equation expressed in the Lorentz gauge (when
integrated with the metric perturbation $h_{ab}$, the first term on
the right-hand side of Eq.~(\ref{linearop}) corresponds to the
linearized Einstein tensor, whereas the last two terms correspond to
$\langle \hat{T}^{(1)}_{ab}[g+h] \rangle$, as follows from
Eq.~(\ref{STexpecvalue})). Furthermore,
$Z^\mathrm{(LO)}_\mathrm{CTP}[J_{ab}^{\Sigma} = 0, J_{cd}^{\Delta}]$
is also the generating functional for the stochastic correlation
functions for the solutions of the Einstein-Langevin equation and,
therefore, the stochastic correlation functions are actually
equivalent to quantum correlation functions for the metric
perturbations.

The solutions of Eq.~(\ref{langevin2}) can be expressed as
\begin{equation}
\Sigma_{ab}(x) = \Sigma_{ab}^{(0)}(x) + \bar{\kappa} (G_\mathrm{ret}
\cdot \xi)_{ab}(x)
\label{solution},
\end{equation}
where $\Sigma_{ab}^{(0)}$ is a solution of the homogeneous equation
$(L \cdot \Sigma)_{ab} = 0$, which coincides with the linearized
semiclassical Einstein equation for the metric perturbations in the
Lorentz gauge, with all the information on the initial conditions, and
$(G_\mathrm{ret})_{abcd}(x,x')$ is the retarded propagator associated
with the integro-differential operator $L_{abcd}(x,x')$ with vanishing
initial conditions on $\mathcal{S}_i$. The CTP generating functional
for a nonvanishing $J_{ab}^{\Sigma}$ can then be written as follows:
\begin{equation}
Z^\mathrm{(LO)}_\mathrm{CTP}[J_{ab}^{\Sigma},J_{cd}^{\Delta}]
= \left\langle e^{i J^{\Delta} \cdot \Sigma^{(0)}} \right\rangle
_{\Sigma_{ab}^{(i)}, \Pi^{cd}_{(i)}}
e^{i\frac{2 \bar{\kappa}}{N} J^{\Delta} \cdot G_\mathrm{ret} \cdot J^{\Sigma}}
e^{-\frac{\bar{\kappa}^2}{2N} J^{\Delta} \cdot G_\mathrm{ret} \cdot \mathcal{N}
\cdot G_\mathrm{ret}^T \cdot J^{\Delta}} \label{ZCTP4},
\end{equation}
where we introduced the notation $(A^T)_{abcd}(x,y) \equiv
A_{cdab}(y,x)$.

It is interesting to consider the particular case of the two-point
correlation function. Functionally differentiating twice with respect
to $J_{ab}^{\Delta}$ and then taking $J_{ab}^{\Delta}$ and
$J_{ab}^{\Sigma}$ equal to zero,  one gets the following result for the
symmetrized quantum correlation function for the metric perturbations:
\begin{equation}
\frac{1}{2}\left\langle \left\{ \hat{h}_{ab}(x), \hat{h}_{cd}(x')
\right\} \right\rangle
= \left\langle \Sigma_{ab}^{(0)}(x) \Sigma_{cd}^{(0)}(x') \right\rangle
_{\Sigma_{ab}^{(i)}, \Pi^{cd}_{(i)}}
+ \frac{\bar{\kappa}^2}{N} \left( G_\mathrm{ret} \cdot \mathcal{N}
\cdot (G_\mathrm{ret})^{T}
\right)_{abcd} (x,x')
\label{correlation}.
\end{equation}
One can see that there are two separate contributions to the two-point
correlation function: the first one is related to the dispersion of
the initial state for the metric perturbations, whereas the second one
is proportional to the noise kernel and accounts for the fluctuations
induced by their interaction with the environment (in this case, the
quantum matter fields). We refer to these two contributions as
\emph{intrinsic} and \emph{induced} fluctuations respectively.
Furthermore, taking into account Eq.~(\ref{ZCTP5}), we see that,
under the aforementioned conditions, the
symmetrized quantum correlation function for the metric perturbations
is equivalent to the stochastic correlation function
obtained in stochastic semiclassical gravity by solving the
Einstein-Langevin equation.

{F}rom the expression for the generating functional in
Eq.~(\ref{ZCTP4}) one can get the remaining two-point quantum
correlation functions to leading order in $1/N$. In particular the
commutator is given by $\langle [\hat{h}_{ab}(x), \hat{h}_{cd}(x')]
\rangle = 2i\kappa (G_\mathrm{ret} (x',x) - G_\mathrm{ret} (x,x'))$,
and by combining the commutator and the anticommutator the rest of
two-point functions can be easily obtained. Moreover, assuming a
Gaussian initial state with vanishing expectation value for the metric
perturbations, the expression for the generating functional in
Eq.~(\ref{ZCTP4}) becomes Gaussian and any other $n$-point quantum
correlation function has a simple expression in terms of the two-point
functions.

\begin{figure}
\includegraphics[height=5cm]{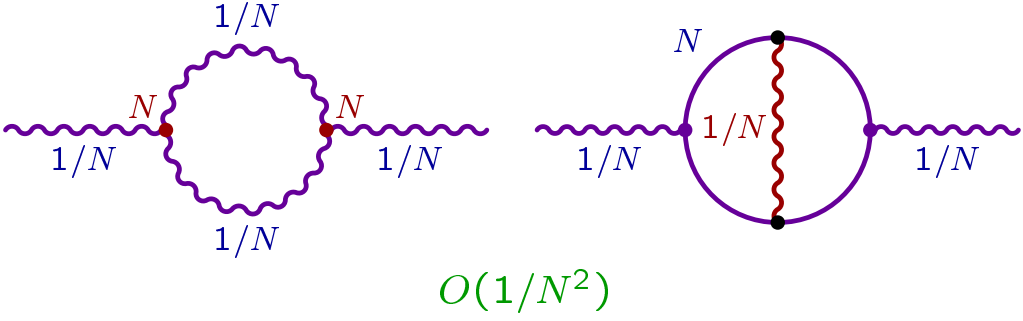}
\caption[fig1]{Two diagrams illustrating the fact that including
either the vertices for the metric perturbations (as in the first
diagram) or terms from the influence functional evaluated beyond the
Gaussian approximation (as in the second diagram) leads to
contributions of higher order in $1/N$. In particular, the two
diagrams shown here give contributions of order $1/N^2$ to the
two-point quantum correlation function for the metric
perturbations. The plain lines represent the CTP propagators for the
matter fields on the background spacetime and the wavy lines
correspond to the CTP propagators for the metric perturbations
obtained by functionally differentiating
$W^\mathrm{(LO)}_\mathrm{CTP}$ twice with respect to the external
currents.}
\label{fig1}
\end{figure}

The exact CTP generating functional is given by
\begin{equation}
Z_\mathrm{CTP}[J_{ab}^{\Sigma},J_{cd}^{\Delta}]
= \exp\left(i S_\mathrm{int}\left[\frac{\delta}{\delta J^{\Sigma}_{ab}},
\frac{\delta}{\delta J^{\Delta}_{cd}}\right]\right)
Z^\mathrm{(LO)}_\mathrm{CTP}[J_{ab}^{\Sigma},J_{cd}^{\Delta}]
\label{exactZCTP},
\end{equation}
where $S_\mathrm{int}[\Sigma_{ab},\Delta_{cd}]$ corresponds to all the
terms in the gravitational action or the exact influence action of
cubic or higher order in the metric perturbations. In order to
consider and evaluate the different contributions to
Eq.~(\ref{exactZCTP}), it is convenient to introduce the corresponding
Feynman rules and diagrams (in the CTP formulation) as follows: each
term in $S_\mathrm{int} \left[ \delta / \delta J^{\Sigma}_{ab}, \delta
/ \delta J^{\Delta}_{cd} \right]$ gives rise to a vertex with the same
number of legs as the total power of the functional derivatives
$\delta / \delta J^{\Sigma}_{ab}$ and $\delta / \delta
J^{\Delta}_{cd}$ appearing in that term, and the CTP propagators
simply correspond to those obtained by functionally differentiating
$W^\mathrm{(LO)}_\mathrm{CTP} = -i \ln Z^\mathrm{(LO)}_\mathrm{CTP}$
with respect to the external currents twice. Expanding in powers of
$1/N$, one can show that all the diagrams representing the
corrections, as given by Eq.~(\ref{exactZCTP}), to the connected part
of the generating functional, $W_\mathrm{CTP} = -i \ln Z_\mathrm{CTP}$,
are of order $1/N^2$ or higher \cite{roura03b}. Therefore, one can
conclude that the leading order contribution to $W_\mathrm{CTP}$ is
entirely given by $W^\mathrm{(LO)}_\mathrm{CTP}$, which is of order
$1/N$ and from which the leading order contribution to all the quantum
correlation functions with an even number of points can be
obtained. Two particular examples showing how the corrections due to
$S_\mathrm{int}[\Sigma_{ab},\Delta_{cd}]$ contribute to the
two-point quantum correlation functions (corresponding to terms in
$W_\mathrm{CTP}$ which are quadratic in the external currents) are
provided in Fig.~\ref{fig1}. The first diagram involves vertices with
three legs associated with cubic terms in the gravitational
action. The second diagram involves a nonlocal vertex with four legs
associated with quartic terms in the influence action. The nonlocal
vertex has been represented by a loop of the matter fields because, if
Feynman diagrams are introduced when evaluating the influence action,
the terms quartic in the metric perturbations giving rise to the second
diagram in Fig.~\ref{fig1} correspond to a loop of matter fields with
four insertions linear in the metric perturbations (there are three other
contributions to the influence action involving terms quartic in the
metric perturbations: one corresponds to a loop of matter fields with
two insertions linear in the metric perturbation and a third insertion
quadratic in the metric perturbation, a second contribution that
corresponds to a loop of matter fields with two insertions quadratic
in the metric perturbation, and a third one corresponding to a loop of
matter fields with an insertion linear in the metric perturbation and
a second insertion cubic in the metric perturbation).

\section{Singular coincidence limit for the noise kernel}
\label{appB}

The noise kernel defined by Eq.~(\ref{noise}) has in general a
singular coincidence limit $x \rightarrow x'$ (in fact, it is still
singular even for $x \neq x'$ when the two points are connected by a
null geodesic), which translates into an ultraviolet divergence when
integrating over momenta in Fourier space, as can be seen from
Eq.~(\ref{noiseMink}). The result is, nevertheless, finite when $x
\neq x'$ (and they are not connected by a null geodesic).  In fact,
even though the noise kernel is not well defined as a tensor-valued
function \footnote{The noise kernel is in general a bitensor, but due
to the homogeneity of Minkowski spacetime and the triviality of the
connection (and the corresponding parallel transport), the noise
kernel becomes in that case a simple tensorial field which depends on
$(x-x')^\mu$.}, it is well defined as a tensor-valued distribution and
yields finite results when integrated with suitable test functions.

Let us consider a specific example to illustrate the points addressed
in this Appendix: a massless conformally coupled scalar field in
Minkowski spacetime. The expression for the noise kernel in spacetime
coordinates, which results from Fourier transforming
Eq.~(\ref{noiseMink}) and is a well-defined distribution, is the
following \cite{martin00}:
\begin{equation}
\mathcal{N}_{\mu\nu\rho\sigma}(x-x') \propto (\partial_\mu\partial_\nu
\partial'_\rho\partial'_\sigma) \,
\mathcal{P}\!f\! \left(\frac{1}{(x-x')^2}\right)^2
= \frac{1}{16}(\partial_\mu\partial_\nu \partial'_\rho\partial'_\sigma)
\Box_x \Box_x' \ln (x-x')^2
\label{distribution},
\end{equation}
where $\mathcal{P}\!f\!$ stands for the Hadamard finite part
prescription, whose precise definition can be found in
Refs.~\cite{schwartz57,zemanian87}. However, the contribution from
the induced fluctuations to the symmetrized two-point correlation
function, which is given by
\begin{equation}
\frac{\bar{\kappa}^2}{N} \left( G_\mathrm{ret} \cdot \mathcal{N}
\cdot \left(G_\mathrm{ret}\right)^{T} \right) (x_1,x_2)
\label{convolution},
\end{equation}
is not necessarily well-defined if the time integral in $\cdot$
involves a finite initial time $t_i$. That is because in that case the
noise kernel is actually convoluted with $G_\mathrm{ret}(x_1,x'_1)
\theta(t'_1-t_i)$, which is not a good test function since it is not
differentiable at $t'_1 = t_i$. The fact that the result for
expression~(\ref{convolution}) is singular for a finite initial time
can actually be seen by using the last equality in
Eq.~(\ref{distribution}), substituting into
expression~(\ref{convolution}) and integrating by parts. The
contributions from the boundary terms at the finite times $t_1$ and
$t_2$ are finite, at least when $t_1 \neq t_2$ (the fact that $t_1
\neq t_2$ may be required to get a finite result is simply indicating
that expression~(\ref{convolution}) is also a distribution). On the
other hand, the boundary terms that correspond to taking $t'_1$ and
$t'_2$ both equal to $t_i$ are divergent. The fact that all the
singular contributions can be concentrated at the initial time seems
to suggest that the origin of the problem may be related to the
initial state that was chosen.

We proceed now to argue that the origin of the singularities described
in the previous paragraph can indeed be traced back to the initial
state that was considered, with the metric perturbations and the matter
fields completely uncorrelated. In order to do that, it will be
useful to discuss an analogous situation for QBM models such as that
described in Appendix~\ref{appA}. In particular, let us consider an
ohmic distribution for the environment frequencies with an ultraviolet
cut-off $\Lambda$, which can be characterized by a spectral density
function such as $I(\omega) = \omega \theta(\Lambda - \omega)$ or
$\omega \exp(-\omega / \Lambda)$ (the details about the particular way
in which the cut-off is implemented are not important here).  We have
an expression analogous to (\ref{convolution}) for the induced
fluctuations (see Eq.~(\ref{correlQBM2}) and
Ref.~\cite{calzetta03a}). If we consider the ground state (thermal
state at zero temperature) as the initial state for the environment,
the noise kernel is given by $N(t,t') = \int d\omega I(\omega) /
\omega \cos \omega (t-t')$. When taking the limit $\Lambda \rightarrow
\infty$ the noise kernel becomes proportional to $\mathcal{P}\!f\!
\left( 1/(t'_1-t'_2)^2 \right)$. To obtain the correlation function we
integrate by parts, as described above, in the expression for the
induced fluctuations, Eq.~(\ref{correlQBM2}), before taking the limit
$\Lambda \rightarrow \infty$. We obtain again a boundary term at the
initial time which diverges as we finally let $\Lambda$ go to
infinity, and we end up with an infinite result for the correlation
function. On the other hand, one can show that the result for the
correlation function of the ground state of the whole system (system
plus environment), including the system-environment interaction, is
finite (the use of Euclidean path integrals is particularly convenient
in this respect)  \cite{grabert88}. This constitutes a clear example of
the fact that initial states in which the system and the environment
are suitably correlated give rise to well-defined correlation
functions.

Alternatively, when taking a completely uncorrelated initial state,
one can still get a finite result for the correlation function by
smoothly switching on the system-environment interaction so that the
boundary term at the initial time which results from the integration
by parts and becomes divergent in the limit $\Lambda \rightarrow
\infty$ actually vanishes. This reveals again that the origin of the
singularity for the correlation function arises because the highest
frequency modes of the environment become correlated with the system
in a timescale of the order of $\Lambda^{-1}$. Such a fact is
supported by the existence of a jolt with a characteristic timescale
$\Lambda^{-1}$ in the diffusion coefficients of the master equation
which becomes singular when $\Lambda \rightarrow \infty$, as was found
in Ref.~\cite{hu92}. In fact, one can show that those states in which
the high frequency modes of the environment and the system are
uncorrelated are unphysical when the environment contains an infinite
number of modes with arbitrarily high frequencies since their energy
becomes infinite as $\Lambda \rightarrow \infty$.

Return now to the gravitational case. There are some situations, such as
the effect of stress tensor fluctuations on the propagation of null
geodesics, in which the appropriate way to deal with
the singular coincidence limit of the noise kernel is by integrating
over some smearing function \cite{borgman03} (in general
smearing just along the spatial directions is not enough: smearing in
time is needed to get a finite result). On the other hand, when
computing the correlation functions for the metric perturbations, the
noise kernel naturally appears integrated with the retarded
propagator. As explained above, the problem still persists at the
initial time, which reflects the unphysical character of the
completely uncorrelated initial state that was employed. Similar to
QBM models, a well-defined result for the correlation
functions can be obtained by considering a properly correlated initial
state, such as that resulting from the use of Euclidean path integrals
that are then analytically continued to Lorentzian time
\cite{hawking01}.  Roughly speaking, this would imply the existence
of an additional term in Eqs.~(\ref{correlation2}) and (\ref{correlation})
due to the existence of correlations between the initial conditions
for the solutions of the Langevin equation and the stochastic source,
which reflect the initial correlations between the system and the
environment.

Alternatively, one can still make sense of the results obtained from
assuming an uncorrelated initial state by smoothly switching on the
interaction between the metric perturbations and the matter fields so
that the high frequency modes can get correlated with the
system. However, in contrast to the QBM case, we have to be careful
with switching on the interaction during a finite period of time since
that would imply that the source of the Einstein-Langevin equation is
not conserved and would be in conflict with the Bianchi identity,
which guarantees the integrability of the equation. Therefore, the
interaction should be turned on adiabatically and asymptotically past
initial conditions should be considered. In fact, in Sec.~\ref{sec5},
where we assumed asymptotic initial conditions and worked mostly in
Fourier space, a finite result for the correlation function was
obtained without the need for explicitly switching on the interaction
adiabatically.  There are, however, situations (for instance, in
cosmology) in which asymptotic initial conditions are not adequate. An
alternative procedure should be considered in those cases.

\section{Runaway solutions and methods to deal with them}
\label{appC}

In this Appendix we will briefly discuss the existence of runaway
solutions in SCG (solutions which grow without bound in timescales
comparable to the Planck time), their counterparts at the quantum
level, and how their connection can be understood in the context of
stochastic gravity. We will also discuss the existing prescriptions
for dealing with this kind of unstable solutions.

\subsection{Runaway solutions in semiclassical gravity}

Let us start by considering the linearized semiclassical Einstein
equation around the Minkowski spacetime. The solutions for the case of
a massless scalar field were first discussed in Ref.~\cite{horowitz80}
and an exhaustive description can be found in Appendix~A of
Ref.~\cite{flanagan96}. Taking Eq.~(\ref{einstein3}) and using a
decomposition for the linearized Einstein tensor analogous to that
introduced in Sec.~\ref{sec6} for the metric perturbation, the
vectorial part is found to vanish \footnote{More precisely,
decomposing the metric perturbation into scalar, vectorial and
tensorial parts, as done in Sec.~\ref{sec6}, and computing the
linearized Einstein tensor, one gets a vanishing result for the
vectorial part of the metric perturbation; the scalar and tensorial
components of the metric perturbation give rise, respectively, to the
scalar and tensorial components of the linearized Einstein tensor.},
whereas the scalar and tensorial contributions satisfy the following
equations:
\begin{eqnarray}
&\left( F_1(p) + 3 p^2 F_2(p) \right) \tilde{G}^{(1)\, \mathrm{(S)}}
_{\mu\nu}(p) = 0 ,&\\
&F_1(p) \tilde{G}^{(1)\, \mathrm{(T)} }_{\mu\nu}(p) = 0 .&
\end{eqnarray}
where $F_1(p)$ and $F_2(p)$ are given by Eqs.~(\ref{F1})-(\ref{F2}),
and $\tilde{G}^{(1) \, \mathrm{(S)}}_{\mu\nu}$ and $\tilde{G}^{(1)\,
\mathrm{(T)}}_{\mu\nu}$ denote, respectively, the scalar and tensorial
parts of the linearized Einstein tensor.  In order to illustrate how
the runaway solutions arise, we will consider the particular example
of a massless and conformally coupled scalar field (see
Ref.~\cite{flanagan96} for the massless case with arbitrary coupling
and Refs.~\cite{martin00,anderson03} for the general massive
case). The previous equations become then
\begin{eqnarray}
&\left( 1 + 12 \kappa \bar{\beta} p^2 \right) \tilde{G}^{(1)\,
\mathrm{(S)}}_{\mu\nu}(p) = 0 \label{scalar}, &\\
&\lim\limits_{\;\epsilon \rightarrow 0^+} \left( 1 + (960 \pi^2)^{-1} \kappa p^2
\ln \left( \frac{-(p^0 + i \epsilon)^2 + \vec{p}^{\,2} }{\mu^2} \right) \right)
\tilde{G}^{(1)\, \mathrm{(T)} }_{\mu\nu}(p) = 0 \label{tensorial}.&
\end{eqnarray}
In addition to the obvious solution $\tilde{G}^{(1)\, \mathrm{(S)}
}_{\mu\nu}(p) = 0$ (the only solution when $\bar{\beta} = 0$), when
$\bar{\beta} > 0$ the solutions for the scalar component exhibit an
oscillatory behavior in spacetime coordinates which corresponds to a
massive scalar field with $m^2 = (12 \kappa |\bar{\beta}|)^{-1}$; for
$\bar{\beta} < 0$ the solutions correspond to a
tachyonic field with $m^2 = - (12 \kappa |\bar{\beta}|)^{-1}$: in
spacetime coordinates they exhibit an exponential behavior in
time --growing or decreasing-- for wavelengths larger than $4 \pi (3
\kappa |\bar{\beta}|)^{1/2}$, and an oscillatory behavior for
wavelengths smaller than $4 \pi (3 \kappa |\bar{\beta}|)^{1/2}$.
On the other hand, the solution
$\tilde{G}^{(1)\, \mathrm{(S)} }_{\mu\nu}(p) = 0$
is completely trivial since any scalar metric perturbation
$\tilde{h}_{\mu\nu}(p)$ giving rise to a vanishing linearized Einstein
tensor can be eliminated by a gauge transformation as explained in
Sec.~\ref{sec6}.

As for the tensorial component, when $\mu \le \mu_\mathrm{crit} =
l_p^{-1} (120\pi)^{1/2} e^{\gamma}$ (or $\lambda \ge
\lambda_\mathrm{crit} = \mu_\mathrm{crit}^{-1}$ in the notation of
Ref.~\cite{flanagan96}) the first factor in Eq.~(\ref{tensorial})
vanishes for four complex values of $p^0$ of the form $\pm \omega$ and
$\pm \omega^*$, where $\omega$ is some complex value, as illustrated
in Fig.~\ref{fig3}. We will consider here the case in which $\mu <
\mu_\mathrm{crit}$; a detailed description of the situation for $\mu
\ge \mu_\mathrm{crit}$ can be found in Appendix~A of
Ref.~\cite{flanagan96}. The two zeros on the upper half of the complex
plane correspond to solutions in spacetime coordinates exponentially
growing in time, whereas the two on the lower half correspond to
solutions exponentially decreasing in time. Strictly speaking, these
solutions only exist in spacetime coordinates, since their Fourier
transform is not well defined. They are commonly referred to as
runaway solutions and for $\mu \sim l_p^{-1}$ they grow exponentially
in timescales comparable to the Planck time.

\subsection{Quantum mechanical systems with higher order time derivatives}

Before proceeding to discuss the situation in stochastic gravity, it
is interesting to make a few remarks about the quantization of higher
derivative theories and the counterparts of the previous classical
instabilities in the quantum context. Let us consider first a free
theory with a structure analogous to that of linearized semiclassical
gravity around Minkowski spacetime without including the nonlocal
terms.  It is characterized by the following Lagrangian, which
corresponds to a harmonic oscillator with a higher derivative term:
\begin{equation}
L(q,\dot{q},\ddot{q}) = \frac{\tau}{2} \ddot{q}^2 + \frac{1}{2}
 \dot{q}^2 - \frac{1}{2} \Omega^2 q^2
\end{equation}
To begin with, one can consider a generalization of the usual
canonical formalism introduced by Ostrogradski to deal with theories
involving higher order derivatives (see, for instance
Refs.~\cite{simon90,urries98}). The theory can then be
quantized following the standard canonical quantization rules. The
corresponding Wigner function (or Wigner functional if a field theory
were considered) can also be introduced. The pathological character of
the theory becomes clear by diagonalizing the Hamiltonian and
realizing that the result corresponds to two independent harmonic
oscillators, but with one of them having a negative sign in the
kinetic term. For $\tau < 0$ the potential term of the harmonic
oscillator with the negative kinetic term is also negative and the
classical solutions do not exhibit instabilities. However, in any case
the configurations for the harmonic oscillator with the negative
kinetic term can have negative energies arbitrarily large in absolute
value. Moreover, the frequency for that oscillator is proportional to
$\tau^{-1/2}$ and diverges as $\tau \rightarrow 0$.  At the quantum
level, such a theory also gives rise to negative eigenvalues of the
Hamiltonian arbitrarily large in absolute value, but can be
alternatively formulated in terms of a Hamiltonian without negative
energies by introducing states with negative norm (commonly referred
to as ghosts) \cite{hawking02}. This fact is often argued in a
qualitative way by pointing out that the propagator of the theory in
Fourier space is proportional to
\begin{equation}
\frac{1}{\omega^2 - \Omega^2} - \frac{1}{\omega^2 + \tau^{-1}} .
\end{equation}
It should also be mentioned that Hawking and Hertog have suggested a
prescription for dealing with that kind of theories which is based on
imposing well-defined boundary conditions in Euclidean time and then
Wick rotating back to Lorentzian time. The results have then a
nonsingular limit $\tau \rightarrow 0$, so that when the higher order
derivative term in the Lagrangian is small, one essentially recovers
the results of the second order theory \cite{hawking02}.

Even though there is a range of parameters ($\tau < 0$) in which the
free theory described above does not exhibit instabilities, they arise
when a nonlinear self-interaction term is added to the Lagrangian. The
reason is that the two Hamiltonian contributions corresponding to a
couple of harmonic oscillators, one with a negative energy spectrum
and the other with a positive one, can have a stable evolution as long
as they are decoupled. However, adding an interaction term couples
them in such a way that one can acquire negative energies arbitrarily
large in absolute value while the other gains large positive energies,
which is the source of instability. In general this is reflected in
the structure of the propagator as a shift of the poles on the real
axis to the complex plane. Hawking \emph{et al.} have argued that
well-behaved results can still be obtained by imposing boundary
conditions which discard solutions which grow unboundedly in time
\cite{hawking01,hawking02}. Those conditions can be implemented by a
suitable choice of the integration contour on the complex plane when
computing the inverse Fourier transform of the propagator, but
causality is violated at small timescales (we will come back to this
point below). Another possibility, when the parameter $\tau$ is small,
is to make use of an order reduction procedure
\cite{landau75,parker93,flanagan96}, which consists of differentiating
the equation of motion with respect to time, substituting back into
the original equation and discarding the terms of higher order in
$\tau$. This procedure can be iterated as many times as necessary to
get a second order equation valid up to the corresponding order of
$\tau$. The usual canonical formalism associated with the second order
equation of motion can then be employed to evolve the Wigner
function. It should be stressed that, although we have considered a
simple model as an illustrating example, the previous methods have
been applied to more involved situations, including SCG
\cite{parker93,flanagan96} and quantum cosmology \cite{hawking01}.

\subsection{Runaway solutions in stochastic gravity}
\label{appC.3}

Let us now address the case of stochastic gravity and see how the
instabilities in SCG and the difficulties in quantizing theories with
higher order derivatives are related. First of all, we recall that in
Appendix~\ref{appE} the counterterms quadratic in the curvature were
ignored and it was implicitly assumed that the Einstein-Langevin
equation was a second order integro-differential equation whose initial
conditions were completely determined by specifying the metric
perturbation and its normal derivative on the initial Cauchy
hypersurface. If the counterterms quadratic in the curvature, which
give rise to higher order derivative terms, are also taken into
account, the generalized canonical formalism referred to above and the
corresponding Wigner functional should be used. In fact, due to the
singular behavior of the nonlocal part of the dissipation kernel at
the initial time, specifying initial conditions at a finite initial
time is an even more delicate matter. In any case, since we have to
consider asymptotic initial conditions to deal with the singular
coincidence limit of the noise kernel, as explained in
Appendix~\ref{appB}, we do not need to be concerned about the problems
associated with finite initial times. Runaway solutions, however,
still exist and some method to deal with them is required. In
particular, when computing two-point correlation functions in the
context of stochastic gravity, the existence of runaway solutions has
implications for both the intrinsic and the induced contributions.

One possible method for dealing with the existence of runaway
solutions is the \emph{order reduction} prescription. As explained
above, the method is based on treating perturbatively the terms
involving higher order derivatives, differentiating the equation
under consideration and substituting back the higher derivative
terms in the original equation keeping only terms up to the
required order in the perturbative parameter. In the case of the
semiclassical Einstein equation, the perturbative parameter
employed is $\hbar$ or, equivalently, the square of the Planck
length $l_p^2 = \kappa / 8 \pi$. If we consider the semiclassical
Einstein equation for linear metric perturbations around
Minkowski spacetime and differentiate twice with respect to the
background covariant derivative, it becomes clear that the second
order derivatives of the Einstein tensor are of order $\kappa$.
Substituting back into the original equation, we get the
following equation up to order $\kappa^2$:
\begin{equation}
G^{(1)}_{ab} [g+h] = 0 + O(\kappa^2)
\label{redeinstein},
\end{equation}
where no effects from the vacuum polarization of the quantum matter
fields are left. Since the linearized semiclassical Einstein equation
coincides with the homogeneous part of the Einstein-Langevin equation,
Eq.~(\ref{redeinstein}) governs the contribution of the intrinsic
fluctuations to the quantum correlation function, which coincides with
that of free gravitons. Similarly, when making use of the order
reduction prescription, the Einstein-Langevin equation becomes
\begin{equation}
G^{(1)}_{ab} [g+h] = \kappa \xi_{ab} + O(\kappa^2)
\label{redlangevin},
\end{equation}
where the stochastic source, whose correlation function only depends
on the background metric and hence does not involve higher order
derivatives of the metric perturbation, is not affected by the order
reduction procedure. Therefore, in contrast to the intrinsic
fluctuations, there will still be a nontrivial contribution to the
induced fluctuations due to the polarization of the quantum matter
fields, but no contribution from the dissipation kernel is left in the
Einstein-Langevin equation. Since all the terms involving higher order
derivatives, which were associated with the dissipation kernel, have
been discarded, an ordinary Wigner functional can be introduced
without any need to consider generalized Ostrogradski
momenta. Furthermore, the absence of the dissipation kernel also
allows the possibility of specifying initial conditions at a finite
initial time as far as the homogeneous solutions (relevant for the
computation of the intrinsic fluctuations) and the retarded propagator
are concerned. Nevertheless, one is still forced to consider
asymptotic initial conditions in order to get a finite result for the
induced fluctuations due to the singular coincidence limit of the
noise kernel, as explained in the previous Appendix.

\begin{figure}[ht]
\includegraphics[height=4cm]{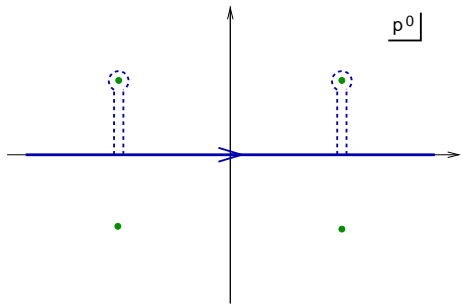}
\caption[fig3]{Representation in the complex plane of the values of
$p^0$ for which the coefficient of the semiclassical Einstein equation
for the tensorial components of the Einstein tensor in Fourier space
vanishes. The case $\mu < \mu_\mathrm{crit}$ and a particular value of
$\vec{p}^{\,2}$ were considered, but the qualitative structure will
remain the same for any other value of $\vec{p}^{\,2}$. The plot also
corresponds to the structure of the poles of the tensorial part of the
propagator $\tilde{D}_{\mu\nu\alpha\beta}(p)$. The solid line
corresponds to the integration contour when Fourier transforming back
to spacetime coordinates which follows from the prescription proposed
by Hawking \emph{et al.} \cite{hawking01} and was chosen in
Ref.~\cite{martin00}. Changing this contour as indicated by the dashed
lines, one obtains a strictly retarded propagator, but it exhibits
exponential instabilities for large positive time differences
associated with the two poles on the upper half of the complex plane.}
\label{fig3}
\end{figure}

Hawking \emph{et al.} have proposed an alternative procedure for
dealing with the runaway solutions \cite{hawking01, hawking02}. Their
method is based on imposing final boundary conditions which discard
those solutions that grow unboundedly in time. Let us first see how
their approach can be applied to the computation of the intrinsic
fluctuations by considering the particular case of a massless and
conformally coupled scalar field. From Eq.~(\ref{tensorial}) and
Fig.~\ref{fig3} one can see that, in addition to the solution
$G^{(1)}_{\mu\nu} = 0$, the solutions of the tensorial part grow or
decrease exponentially in time. The exponentially growing solutions
are discarded when the final boundary condition is imposed, and the
contributions from the exponentially decreasing ones also vanish if
regular initial conditions are specified at an asymptotic initial
time. On the other hand, from Eq.~(\ref{scalar}) one can see that the
situation is analogous for the solutions of the scalar part when
$\bar{\beta} < 0$. For $\bar{\beta} > 0$ the solutions are oscillatory
and, hence, are not discarded when the final boundary condition
is imposed (in contrast to the situation where the order reduction
prescription is used).

Let us now apply the previous approach to the computation of the
induced fluctuations. When considering asymptotic initial conditions,
the relevant propagator for expressing the linearized Einstein tensor
in terms of the stochastic source, can be obtained by inverting
$F_{\mu\nu\alpha\beta}(p)$ in Eq.~(\ref{einstein3}). The resulting
propagator, $\tilde{D}_{\mu\nu\alpha\beta}(p)$, exhibits a number of
poles in the complex plane, as illustrated in Fig.~\ref{fig3}. The
expression for the retarded propagator in spacetime coordinates
corresponds to choosing the integration path represented by the dashed
line in Fig.~\ref{fig3} when Fourier transforming back from momentum
space. It exhibits the appropriate causal behavior:
$D_{\mu\nu\alpha\beta}(x-y) = 0$ for $t_x < t_y$, as can be seen by
closing the integration contour on the upper half of the complex
plane. However, for $t_x > t_y$ it increases exponentially in time due
to the contributions from the two poles on the upper half of the
complex plane when closing the path on the lower half. Imposing the
final boundary conditions which discard solutions growing unboundedly
in time is equivalent to taking a different integration path: that
represented by a solid line in Fig.~\ref{fig3}. The resulting
propagator does not exhibit exponential instabilities, but gives rise
to causality violations since $D_{\mu\nu\alpha\beta}(x-y) \neq 0$ for
$t_x < t_y$ (the characteristic timescale of these causality
violations is of order $\sqrt{N}l_p$).  This propagator is the only
one which has a well-defined Fourier transform. It was employed in
Ref.~\cite{martin00}, where it was argued that any other propagator
should yield an equivalent result for the correlation function
obtained by solving the Einstein-Langevin equation. This argument is
certainly true for propagators with a well-defined Fourier
transform. However, the existence of poles off the real axis gives
rise to propagators in spacetime coordinates (they do not have a
well-defined Fourier transform because of the exponentially growing or
decreasing contributions) which yield inequivalent results for the
correlation function. Since this choice for the propagator was made,
the results obtained in Ref.~\cite{martin00} correspond to those that
would follow when employing the procedure proposed by Hawking \emph{et
al.}. In fact, Hawking \emph{et al.} applied their method to quantum
propagators, but, as we have described, it can also be used when
solving the semiclassical Einstein equation and the Einstein-Langevin
equation. The stochastic correlation functions obtained are then
equivalent to the quantum correlation functions (CTP propagators)
which would result from the application of the prescription.

\subsection{Estimates of radiative corrections for a single matter
field and a large number of them}

SCG is expected to provide reliable results as long as the
characteristic lengthscales under consideration are much larger than
the Planck length $l_p$ \cite{flanagan96}. This can be qualitatively
argued by estimating the magnitude of the different contributions to
the effective action (considering the relevant Feynman diagrams and
using dimensional arguments): the Einstein-Hilbert term and the
radiative quantum corrections. The Einstein-Hilbert term is of order
$l_p^{-2} R$ (the characteristic curvature $R$ is simply given by
$L^{-2}$, where $L$ is the characteristic lengthscale of our problem);
the vacuum polarization terms involving loops of matter fields are of
order $R^2$; and higher loop corrections involving internal graviton
propagators are of order $l_p^2 R^3$ or higher. Thus, we see that the
higher order corrections not included in SCG are negligible provided
that $L \gg l_p$. In that regime, however, the vacuum polarization
terms only yield a small correction to the Einstein-Hilbert term and
any classical gravitational source which were present. The
justification of the order reduction prescription is actually based on
this fact. Therefore, significant effects from the vacuum polarization
of the matter fields are only expected when their small corrections
accumulate in time, as would be the case, for instance, for an
evaporating macroscopic black hole all the way before reaching
Planckian scales.

The previous estimates for the different terms in the effective action
change in a remarkable way when a large number of fields, $N$, is
considered \footnote{The actual physical Planck length $l_p$ is
considered, not the rescaled one, $\sqrt{\kappa / 8 \pi}$, which is
related to $l_p$ by $8 \pi l_p^2 = \kappa = \bar{\kappa} / N$.}. The
vacuum polarization terms involving loops of matter become of order $N
R^2$ and, similarly, the higher loop corrections involving internal
graviton propagators are of order $N l_p^2 R^3$ or higher (the
contributions corresponding to one and two graviton loops are,
respectively, of order $R^2$ and $l_p^2 R^3$, but are negligible as
compared to those from matter loops when $N$ is large). There is then
a regime in which the vacuum polarization of the matter fields and the
Einstein-Hilbert term are comparable when $L \sim \sqrt{N}l_p$. On the
other hand, the higher loop corrections will still be much smaller if
$L \gg l_p$. Both conditions are compatible provided that the number
of fields, $N$, is very large. This is, in fact, the kind of situation
considered in trace anomaly driven inflationary models
\cite{hawking01}, such as that originally proposed by Starobinsky
\cite{starobinsky80}, where the exponential inflation is driven by a
large number of massless conformal fields. The order reduction
prescription would completely discard the effect from the vacuum
polarization of the matter fields even though it is comparable to the
Einstein-Hilbert term. In contrast, the procedure proposed by Hawking
\emph{et al.} keeps the contribution from the matter fields.

We conclude this Appendix by mentioning that it has been pointed out
that a similar kind of instability, which is closely connected to the
existence of the Landau pole, is also present in scalar QED (as well
as ordinary QED) \cite{horowitz81,hartle81,jordan87a}. Nevertheless, a
number of nonperturbative studies on the evolution of the expectation
value of the electromagnetic field using a large $N$ expansion have
been carried out. In fact, it was suggested in Ref.~\cite{cooper94}
that by introducing a finite 3-momentum cut-off and considering a
running coupling constant small enough at low energies, the problem
with the Landau pole could be circumvented (at least from a practical
point of view). Yet it seems unlikely that a similar procedure could
work for the gravitational case due to the existence of higher
derivatives. Moreover, introducing a 3-momentum cut-off would break
general covariance and that would pose serious difficulties when
implementing a consistent and natural renormalization scheme in
general curved spacetimes.


\end{document}